\documentclass[10pt]{article}
\usepackage{amsmath}
\usepackage{amssymb}
\usepackage{graphicx}
\usepackage{cite}
\usepackage{color}

\numberwithin{equation}{section}

\newcommand{\til}[1]{\tilde{#1}}
\newcommand{\vev}[1]{\left\langle{#1}\right\rangle}
\newcommand{\del}{\partial}
\newcommand{\csum}{\hat f}
\begin{document}
\baselineskip=18pt
\begin{titlepage}
\begin{flushright}
{\small KYUSHU-HET-176}
\end{flushright}
\begin{center}
\vspace*{11mm}

{\large\bf%
The Standard Model Gauge Symmetry from Higher-Rank Unified Groups in Grand Gauge-Higgs Unification Models%
}\vspace*{8mm}

Kentaro Kojima$^{a,}$\footnote{E-mail: kojima@artsci.kyushu-u.ac.jp}, 
Kazunori Takenaga$^{b,}$\footnote{E-mail: takenaga@kumamoto-hsu.ac.jp}, 
and 
Toshifumi Yamashita$^{c,}$\footnote{E-mail: tyamashi@aichi-med-u.ac.jp}%
\vspace*{5mm}

{\it $^a$ Faculty of Arts and Science, Kyushu University, 
  Fukuoka 819-0395, Japan\\
$^b$ Faculty of Health Science, Kumamoto Health Science University, 
 Izumi-machi, Kumamoto 861-5598, Japan\\
$^c$ Department of Physics, Aichi Medical University, Nagakute, 480-1195, Japan%
}

\vspace*{3mm}

\end{center}
\vspace*{5mm}

\begin{abstract}\noindent%
We study grand unified models in the five-dimensional space-time where
the extra dimension is compactified on $S^1/\mathbb{Z}_2$. The
spontaneous breaking of unified gauge symmetries is achieved via
vacuum expectation values of the extra-dimensional components of gauge
fields. We derive one-loop effective potentials for the zero modes of
the gauge fields in $SU(7)$, $SU(8)$, $SO(10)$, and $E_6$ models. 
In each model, the rank of the residual gauge symmetry that respects the
boundary condition imposed at the orbifold fixed points is higher 
than that of the standard model. We verify that the residual symmetry 
is broken to the standard model gauge symmetry at the global minima of 
the effective potential for certain sets of bulk fermion fields in each model. 
\end{abstract}
\end{titlepage}
\newpage

\tableofcontents 
\vspace{1cm}
\section{Introduction} 
\label{sec:intro}
For the past several decades, grand unification of the standard model
gauge symmetry at a high-energy regime has been considered to be an
attractive idea as physics beyond the standard model, since the
unification helps us to understand unrevealed features involved in the
standard model such as the charge quantization and the anomaly
cancellation. In addition to the minimal grand unified theory (GUT)
based on $SU(5)$~\cite{GG}, there are well known GUT models based on,
for instance, $SU(4)\times SU(2)_L\times SU(2)_R$~\cite{PS},
$SO(10)$~\cite{so10}, and $E_6$~\cite{e6}. A common feature shared
among various GUT models is that some symmetry breaking mechanism is
required to obtain the standard model gauge symmetry
$G_{\rm SM}=SU(3)_C\times SU(2)_L\times U(1)_Y$ at a low-energy
regime.  A standard prescription for the symmetry breaking in GUT
models is to involve elementary Higgs scalars that develop vacuum
expectation values (VEVs) and the VEVs lead to desired breaking
patterns of the unified symmetries. This mechanism is an analogous to
the electroweak symmetry breaking by the Higgs scalar in the standard
model.

Besides the Higgs mechanism, if compactified extra dimensions are
concealed in our Universe, another way of the spontaneous symmetry
breaking becomes possible, namely the Hosotani
mechanism~\cite{manton,fair,hosotani}. In models with the Hosotani
mechanism, the extra-dimensional components of the gauge fields
effectively behave as ``Higgs'' scalars at low energy, and dynamics of
the gauge fields reflects degrees of freedom of Wilson line
phases. Although gauge invariance forbids the tree-level potential for
the phases at the classical level, non-trivial VEVs of the phases are
naturally emerged and spontaneous symmetry breaking is achieved when
quantum corrections are involved~\cite{hosotani}.  Advantages of the
Hosotani mechanism are predictivity and finiteness of the ``Higgs''
potential and masses~\cite{finite}, even though the potential arises
from loop corrections. Hence, as a solution to the hierarchy problem
in the standard model, the gauge-Higgs unification models has been
widely investigated~\cite{hewsb,edgut}.  In these models, the zero
modes of the extra-dimensional component of gauge fields are
identified to the Higgs doublet in the standard model.

Recently, we have been focusing on application of the Hosotani
mechanism to unified gauge symmetry breaking on an orbifold
compactification $S^1/\mathbb{Z}_2$, which enable us to incorporate
chiral fermions in five-dimensional
models~\cite{gghusu5,gghusu5susy}. In this case, the zero mode of the
extra-dimensional gauge field, which have even parities under a
boundary condition defined at the boundaries of the orbifold, plays a
role of the Higgs field whose VEV breaks unified gauge symmetries into
the standard model one. We refer to this scenario as grand gauge-Higgs
unification (gGHU). Note that gGHU is different from the orbifold GUT
models where boundary conditions directly break GUT symmetry into the
standard model gauge symmetry~\cite{orbifoldgut}.

In models with $S^1/\mathbb{Z}_2$, the orbifold parities of the
extra-dimensional component of the gauge field is opposite to those of
the four-dimensional vector counterpart. Consequently, massless zero
modes appearing from the extra-dimensional components tend not to
belong to the adjoint representation of unbroken symmetries, though
adjoint Higgs fields are often utilized in ordinary four-dimensional
GUT models~\cite{GG}. This situation leads to severe constraints on
construction of gGHU models. We have shown that the difficulty is
evaded and the adjoint ``Higgs'' field in the gGHU model is
obtained~\cite{gghusu5} with the diagonal embedding
method~\cite{dembed}, which is known in the context of string
theory. The doublet-triplet splitting problem and several
phenomenological aspects have been studied in the supersymmetric
version of the model~\cite{gghusu5susy}.

In this work, we focus on another way to construct phenomenologically
viable gGHU models where ``Higgs'' fields originated from the
extra-dimensional gauge field transform as non-adjoint representations
of unbroken symmetries. In this case, spontaneous breaking of unified
symmetries triggered by the ``Higgs'' fields generally involves rank
reductions. As concrete examples, we examine the four models based on
the unified symmetries $SU(7)$, $SU(8)$, $SO(10)$, and $E_6$. In each
model, we derive the one-loop effective potential, which depends on
the matter content, for the zero mode of the gauge field. In these
models, accordingly, vacuum structure of the potential and symmetry
breaking pattern are determined by bulk field contents. We show that
the standard model gauge symmetry is achieved at a low-energy regime
for certain sets of bulk fermion fields in each model.

The paper is organized as follows. In Sec.~2, we give a general setup
of five-dimensional gauge theories with a compactified dimension on
$S^1/\mathbb{Z}_2$. A calculation method of the one-loop effective
potential for Wilson line phases is briefly summarized. In Sec.~3, as
illustrative examples of gGHU, we discuss three models where each
unified symmetry is $SU(7)$, $SU(8)$, or $SO(10)$. The one-loop
effective potential in each model is derived. In Sec.~4, the $E_6$
gGHU model is studied and the one-loop effective potential is
examined. Vacuum structure of each effective potential is studied in
Sec.~5.  We finally summarize our discussions in Sec.~6. The
appendices are devoted to show detailed calculations required for the
discussion in Sec.~4.

\bigskip
\section{General setup} 
\label{sec:general}
We consider five-dimensional gauge theories on
$M^4\times S^1/\mathbb{Z}_2$, where one of the spatial dimensions is
compactified on the orbifold $S^1/\mathbb{Z}_2$ and $M^4$ is the
four-dimensional Minkowski space-time. On the compact space $S^1$,
which has the radius $R$, the fifth-dimensional coordinate denoted by
$y$ is identified with $y+2\pi R$ by the translation.  On the orbifold
$S^1/\mathbb{Z}_2$, in addition to the translation, there is another
identification $y\sim -y$, which is induced by the orbifold parity
transformation.  In the $S^1/\mathbb{Z}_2$ orbifold theories, the
combination of the translation and the parity defines another orbifold
parity transformation that leads to the identification
$(\pi R+y)\sim (\pi R-y)$. There exists the gauge field $A_M(x,y)$
that has the four-dimensional part $A_{\mu}$ $(\mu=0,1,2,3)$ and the
extra-dimensional part $A_5$. The gauge field is also denoted by
$A_M=A_M^aT^a$, where $T^a$ is the generator of the gauge symmetry $G$
of the theory.

The above two orbifold parity transformations act on the gauge field
in such a way that the Lagrangian is invariant. Let us define the
parity operators $\hat P_{0,1}$ around $y=0$
\begin{align}
\label{bcsgen1}
  A_\mu(x,-y)&=\hat P_0A_\mu(x,y)=P_0A_\mu(x,y)P_0^\dag, \\
 A_5(x,-y)&=\hat P_0A_5(x,y)=-P_0A_5(x,y)P_0^\dag, 
\end{align}
and around $y=\pi R$
\begin{align}
\label{bcsgen21}
  A_\mu(x,\pi R-y)&=\hat P_1A_\mu(x,\pi R+y)= P_1A_\mu(x,\pi R+y)P_1^\dag,\\
 A_5(x,\pi R-y)&=\hat P_1A_5(x,\pi R+y)=-P_1A_5(x,\pi R+y)P_1^\dag, 
\label{bcsgen2}
\end{align}
where $P_0$ and $P_1$ are matrices in a representation space of $G$.
The orbifold parities and the matrices are called boundary conditions.
Note that the translation from $y$ to $y+2\pi R$ is induced by the
operator $\hat P_1\hat P_0$, whose eigenvalue corresponds to the
periodicity, for each field.

One can introduce Dirac fermions $\psi(x,y)$; they obey the parity
transformations as
\begin{align}
\label{diracbc}
  \psi(x,-y)&=\hat P_0\psi(x,y)=\eta_0 T_\psi[P_0]\gamma^5\psi(x,y), \\
 \psi(x,\pi R-y)&=\hat P_1\psi(x,\pi R+y)=\eta_1T_\psi[P_1]\gamma^5\psi(x,\pi R+y),
\end{align}
where the matrix $T_\psi[P_{0,1}]$ acts on the field in its
representation space and corresponds to $P_{0,1}$ in
Eqs.~\eqref{bcsgen1}--\eqref{bcsgen2}. The parameters $\eta_0$ and
$\eta_1$ can be chosen as $1$ or $-1$ for each fermion. We also use
the notation $\til\eta=\eta_0 \eta_1$, which is related to the
periodicity of each field.

At an energy regime below $1/R$, the theories are well described by
four-dimensional effective pictures where $A_\mu$ and $A_5$ behave as
a four-dimensional vector field and a scalar field, respectively. Once
the boundary condition in Eqs.~\eqref{bcsgen1}--\eqref{bcsgen2} is
specified, we readily see that there exist sets of the generators
$\{T^\alpha\}\subset \{T^a\}$ and
$\{T^{\tilde \alpha}\}\subset \{T^a\}$ that satisfy
$[T^\alpha,P_0]=[T^\alpha,P_1]=0$ and
$\{T^{\tilde \alpha},P_0\}=\{T^{\tilde \alpha},P_1\}=0$,
respectively. The corresponding component fields $A_\mu^\alpha$ and
$A_5^{\tilde \alpha}$ have even parities of both $\hat P_0$ and
$\hat P_1$.  Hence, the Kaluza-Klein (KK) decompositions of
$A_\mu^\alpha$ and $A_5^{\tilde \alpha}$ have massless zero modes. The
set of generators $\{T^\alpha\}$ corresponds to a subgroup of $G$,
which we call the residual gauge symmetry $H$. The massless ``scalar''
zero modes of $A_5^{\tilde \alpha}$ parametrize Wilson line phase
degrees of freedom and can develop non-trivial VEVs
$\langle{A_5^{\tilde \alpha}}\rangle$. As a result, the residual gauge
symmetry $H$ is further broken by the VEVs.

The gauge symmetry forbids tree-level potential for the zero mode of
$A_5$. Therefore, a minimum of the potential, namely vacuum structure
of the theory, is determined by quantum effects. The effective
potential can be derived by the functional integral over fields in the
theory. One can treat $\langle{A_5^{\tilde \alpha}}\rangle$ as
classical backgrounds and substitute
$A_5^{\tilde \alpha}\to A_5^{\tilde \alpha}+\langle{A_5^{\tilde
    \alpha}}\rangle$ in the Lagrangian. Then quadratic terms of the
gauge field in the Lagrangian are written as
\begin{align}
  \label{laggauge}
  {\cal L}_{\rm gauge} 
  &={1\over 2}\eta^{MN}A_M^a  \left[\square \delta^{ac}-(D_y^{(0)})^{ab}(D_y^{(0)})^{bc}\right]A_N^c,\qquad 
    (D_y^{(0)})^{ab}
    =[\del_y\delta^{ab}-ig\langle{A_5^{\tilde \alpha}}\rangle
{\rm ad}(T^{\tilde \alpha})^{ab}
],
\end{align}
where we adopt the Feynman gauge and use
$\eta^{MN}={\rm diag}(1,-1,-1,-1,-1)$ and
$\square = \del_\mu\del^\mu$.  We denote the five-dimensional gauge
coupling constant by $g$ and the generators in the adjoint
representation by ${\rm ad}(T^{\tilde \alpha})^{ab}$ . The functional
integral over the fluctuations $A_M^a$ leads to contributions to the
effective potential for $\langle{A_5^{\tilde \alpha}}\rangle$.

Notice that the contributions to the effective potential are
determined by eigenvalues of the generators $\{T^{\tilde \alpha}\}$,
which are accompanied by the zero modes
$\langle{A_5^{\tilde \alpha}}\rangle$, in the covariant derivative
$D_y^{(0)}$. The generators are regarded as the charge operators of
$U(1)$ subgroups of $G$. Thus once the $U(1)$ charges of the fields in
the functional integral are known, then the contributions to the
effective potential are obtained. In addition, each $U(1)$ generator
is also regarded as the Cartan generator of an $SU(2)$ subgroup of
$G$.  Therefore, we can easily understand the $U(1)$ charges of the
fields from the spin eigenvalues of the $SU(2)$ subgroups accompanied
by the zero modes $\langle{A_5^{\tilde \alpha}}\rangle$~\cite{potfor}.
We use this procedure for deriving the contributions to the effective
potential.

In the following sections, we study several gGHU models that lead to
the standard model gauge symmetry $G_{\rm SM}$ at a low-energy regime.
The gauge symmetry $G$ in gGHU models is broken via boundary
conditions and non-vanishing VEVs $\vev{A_5}$. We also study a model
with gauge symmetry breaking induced by localized anomalies at a
boundary.

\bigskip
\section{Simple examples of gGHU models} 
In this section, we examine three models based on the unified
symmetries $SU(7)$, $SU(8)$, and $SO(10)$. These models are simple and
intuitive examples of the gGHU models that lead to $G_{\rm SM}$ as a
result of boundary conditions and the Hosotani mechanism. The study in
this section helps us to understand the $E_6$ model, which will be
studied in the next section.

We first give a brief explanation for symmetry breaking patterns in
the models studied in this section. In the $SU(N)$ $(N=7,8)$ model, we
adopt the boundary condition that leads to the residual symmetry
$H=SU(5)\times SU(N-5)\times U(1)$, under which the zero mode of $A_5$
transforms as the bi-fundamental representation. Non-zero VEVs of the
Wilson line phases can lead to the spontaneous symmetry breaking
$H\to G_{\rm SM}$. The gauge symmetry breaking of $H$ is similar to
that in the product-group unification~\cite{pgu}, though in which the
construction of the $U(1)_Y$ hypercharge generator is different from
our model.  In the $SO(10)$ model, the residual symmetry is
$H=SU(5)\times U(1)$ and the zero mode of $A_5$ behaves as the
10-dimensional anti-symmetric representation of $SU(5)$. A non-zero
VEV of the 10-dimensional field induces the spontaneous symmetry
breaking $SU(5)\times U(1)\to G_{\rm SM}$. The symmetry breaking
pattern is similar to the flipped $SU(5)$ models~\cite{flipgut},
although there is a difference between the constructions of the
$U(1)_Y$ hypercharge generator in our model and the flipped $SU(5)$
models.

In the following discussions, we give explicit forms of the boundary
condition and the Wilson line phase in each model. The effective
potential for the zero mode of $A_5$ is derived with the calculation
methods in Ref.~\cite{potfor}. The vacuum structures of the effective
potentials will be analyzed in Sec.~\ref{vacan}.

\subsection{The $SU(7)$ and $SU(8)$ models} 
\label{sec:su7}
At first we study the gGHU model with the $SU(7)$ unified symmetry. We choose 
the boundary condition that is defined by the parity matrices: 
\begin{align}
  P_0&=P_1={\rm diag}(1,1,1,1,1,-1,-1). 
\end{align}
This boundary condition implies that the gauge field has the
following eigenvalues of the parity operators:
\begin{align}
  (\hat P_0,\hat P_1)\cdot (A_\mu)&=\left(
    \begin{array}{c|c}
(+,+)\cdot (A_\mu)_{5\times 5}&(-,-)\cdot (A_\mu)_{5\times 2}\\\cline{1-2}
(-,-)\cdot (A_\mu)_{2\times 5}&(+,+)\cdot (A_\mu)_{2\times 2}
    \end{array}
\right), 
 \\
  (\hat P_0,\hat P_1)\cdot (A_5)&=\left(
    \begin{array}{c|c}
(-,-)\cdot (A_5)_{5\times 5}&(+,+)\cdot (A_5)_{5\times 2}\\\cline{1-2}
(+,+)\cdot (A_5)_{2\times 5}&(-,-)\cdot (A_5)_{2\times 2}
    \end{array}
\right), 
\label{a5bc}
\end{align}
where the subscripts in the right-hand sides imply $n\times m$
submatrix in the $SU(7)$ representation space. The zero mode of
$A_\mu$ appears as the adjoint representation of the residual gauge
symmetry $H=SU(5)\times SU(2)\times U(1)$. Meanwhile, $A_5$ has the
zero mode that transforms as the bi-fundamental representation under the
$SU(5)\times SU(2)$ symmetry.

Without loss of generality, the residual symmetry
$SU(5)\times SU(2)\times U(1)$ allows us to simplify the form of the
VEVs of the $A_5$ zero mode as
\begin{align}
\label{su7a5para}
  \vev{A_5}&=
{1\over 2gR}
\left(
\begin{array}{c|c}
{\bf 0}_{5\times 5}&\Theta^a\\\cline{1-2}
(\Theta^a)^\dag&{\bf 0}_{2\times 2}
\end{array}
\right), \qquad 
\Theta^a=
     \begin{pmatrix}
  a_1&0\\
  0&a_2\\
  0&0\\
  0&0\\
  0&0
     \end{pmatrix}, 
\end{align}
where $a_1$ and $a_2$ are real parameters. The gauge symmetry
forbids the tree-level potential for $a_1$ and $a_2$.  If the effective
potential, which is induced by quantum corrections, has the global minima
where $a_1=a_2\neq 0$ (mod 1) is realized, then
$SU(5)\times SU(2)\times U(1)$ is spontaneously broken down to
$G_{\rm SM}$ at a vacuum.\footnote{
Since the parameter $a_i$ $(i=1,2)$ has the phase property, 
the cases with $a_i=0$ and $a_i=2$ are physically equivalent. 
Among the vacua $a_1=a_2\neq 0$ (mod 2), there is 
a special one with $a_1=a_2=1$, where the rank is preserved under the 
spontaneous breaking of the symmetry and 
$SU(3)\times SU(2)\times SU(2)\times U(1)\times U(1)$ 
appears as the low-energy symmetry.}

We start to derive the effective potential. The parametrization of
VEVs in Eq.~\eqref{su7a5para} suggests that the $SU(7)$ generator that
corresponds to $a_1$ $(a_2)$ can be seen as the Cartan generator
of $SU(2)_{16}$ $(SU(2)_{27})$, where $SU(2)_{ij}$ induces mixing
between the $i$-th and $j$-th components of the $SU(7)$ fundamental
representation.  This is an important point to understand eigenvalues
of $D_y^{(0)}$ in the Lagrangian~\eqref{laggauge} and contributions to
the effective potential, as discussed in Sec.~\ref{sec:general}.

In order to derive the effective potential, we consider the
decomposition of the fundamental representation of $SU(7)$ into
representations of $(SU(3)\times SU(2)_{16}\times SU(2)_{27})$ as
\begin{align}
  7  &\to \{(3,1,1)+(1,2,1)\}+\{(1,1,2)\}, 
\end{align}
where representations in each curly bracket compose an irreducible
representation of $SU(5)\times SU(2)_{27}$.  One can see that the
$SU(7)$ fundamental representation involves two doublets under 
$SU(2)_{16}\times SU(2)_{27}$ as
\begin{align}
7\ni 1\times [(2,1)+(1,2)], 
\end{align}
where the right-hand side indicates representations of $(SU(2)_{16},SU(2)_{27})$. 
Similarly, the anti-symmetric 21-dimensional, 
the symmetric 28-dimensional, and the adjoint 
representations are decomposed as follows:
\begin{align}
  21  \to& \{(\bar 3,1,1)+(3,2,1)+(1,1,1)\}+\{(3,1,2)+(1,2,2)\}+\{(1,1,1)\},\\
  28  \to& \{(6,1,1)+(3,2,1)+(1,3,1)\}+\{(3,1,2)+(1,2,2)\}+\{(1,1,3)\},\\
\notag
  48  \to& \{(8,1,1)+(1,3,1)+(1,1,1)+(3,2,1)+(\bar 3,2,1)\}+\{(1,1,3)\}
+\{(1,1,1)\}\\
& +\{(3,1,2)+(1,2,2)\}+\{(\bar 3,1,2)+(1,2,2)\},
\end{align}
in terms of $(SU(3),SU(2)_{16},SU(2)_{27})$. One finds that the
irreducible representations of $SU(7)$ involve the following
$(SU(2)_{16},SU(2)_{27})$ representations:
\begin{align}
    21&\ni 3\times [(2,1)+(1,2)] + 1\times (2,2), \\
    28&\ni 3\times [(2,1)+(1,2)] + 1\times (2,2)+1\times[(3,1)+(1,3)], \\
    48&\ni 6\times [(2,1)+(1,2)] + 2\times (2,2)+1\times[(3,1)+(1,3)], 
\end{align}
and the others are singlets. 

From the above decompositions, one can understand that how the $SU(7)$
representations transform under $U(1)$ generators accompanied by the
parameters $a_1$ and $a_2$. Then the eigenvalues of the covariant
derivative in Eq.~\eqref{laggauge} and the contributions to the
effective potential are easily evaluated.  We denote the contribution
to the effective potential from a bosonic degree of freedom of
the $R$-dimensional $SU(7)$ representation as $F_7^{R}(a_i,\delta)$, where
$\delta=0$ $(-1)$ for the bulk fermion fields of $\til \eta=1$
$(-1)$. For $R=7$, $21$, $28$, and $48$, the contributions are written
as follows:
\begin{align}\label{su7pot7}
  F_7^{7}(a_i, \delta)=&
{-C\over 2}\left[\csum(a_1+\delta)+\csum(a_2+\delta)\right]
=
{-C\over 2}\sum_{i=1}^2\csum(a_i+\delta),\\
\notag  F_7^{21}(a_i, \delta)=&
{-C\over 2}\left[
3\csum(a_1+\delta)+3\csum(a_2+\delta)
+\csum(a_1+a_2+\delta)+\csum(a_1-a_2+\delta)\right]\\
=&
{-C\over 2}
\left[3 \sum_{i=1}^2\csum(a_i+\delta)
+\sum_{1\leq i<j}^2\left\{
\csum(a_i+a_j+\delta)+\csum(a_i-a_j+\delta)\right\}\right]
,\\
\notag  F_7^{28}(a_i, \delta)=&
{-C\over 2}\left[
3\csum(a_1+\delta)+3\csum(a_2+\delta)
+\csum(a_1+a_2+\delta)+\csum(a_1-a_2+\delta)
+\csum(2a_1+\delta)+\csum(2a_2+\delta)
\right]\\
=&
{-C\over 2}
\left[3 \sum_{i=1}^2\csum(a_i+\delta)
+\sum_{1\leq i\leq j}^2\left\{
\csum(a_i+a_j+\delta)+\csum(a_i-a_j+\delta)\right\}\right]
+({\rm constants})
,\\
\notag  F_7^{48}(a_i, \delta)=&
{-C\over 2}\left[
6\csum(a_1+\delta)+6\csum(a_2+\delta)
+2\csum(a_1+a_2+\delta)+2\csum(a_1-a_2+\delta)
+\csum(2a_1+\delta)+\csum(2a_2+\delta)
\right]\\
=&
{-C\over 2}
\left[6 \sum_{i=1}^2\csum(a_i+\delta)
+\sum_{i,j=1}^2\left\{
\csum(a_i+a_j+\delta)+\csum(a_i-a_j+\delta)\right\}\right]
+({\rm constants}), 
\label{su7pot48}
\end{align}
where $C=3/(64\pi^7R^5)$ and we use
\begin{align}
\csum(x+\delta)  &=\sum_{w=1}^{\infty}{\cos\left(\pi w(x+\delta)\right)\over w^5}.
\end{align}
In the above expressions, we denote $a_i$-independent terms as
constants, which have no effect on symmetry breaking patterns and are
discarded in the following discussions. One can confirm that the
results in Eqs.~\eqref{su7pot7}--\eqref{su7pot48} are coincide with
the potential in Ref.~\cite{potfor}.

We specify the matter content for fermion fields in the model as 
\begin{align}
  {\mathcal N}_7\equiv (n_{7}^{(+)},n_{7}^{(-)},n_{21}^{(+)},n_{21}^{(-)},
n_{28}^{(+)},n_{28}^{(-)},n_{48}^{(+)},n_{48}^{(-)}), 
\end{align}
where $n_R^{(\pm)}$ stands for the number of bulk fermion fields that
belong to the $R$-dimensional representations and have 
$\til\eta=\pm 1$.  The one-loop effective potential is written as follows:
\begin{align}\notag
  V_{7}(a_i, {\mathcal N}_7)
=&3F_7^{48}(a_i,0)
-4\bigg[
n_{7}^{(+)} F_7^{7}(a_i,0)
+n_{7}^{(-)} F_7^{7}(a_i,1)
+n_{21}^{(+)} F_7^{21}(a_i,0)
+n_{21}^{(-)} F_7^{21}(a_i,1)\\
&+n_{28}^{(+)} F_7^{28}(a_i,0)
+n_{28}^{(-)} F_7^{28}(a_i,1)
+n_{48}^{(+)} F_7^{48}(a_i,0)
+n_{48}^{(-)} F_7^{48}(a_i,1)\bigg].
\label{su7pot}
\end{align}

Next, let us start to study the gGHU model with the $SU(8)$ unified
symmetry.  We assume the following boundary condition:
\begin{align}
  (\hat P_0,\hat P_1)\cdot (A_\mu)&=\left(
    \begin{array}{c|c}
(+,+)\cdot (A_\mu)_{5\times 5}&(-,-)\cdot (A_\mu)_{5\times 3}\\\cline{1-2}
(-,-)\cdot (A_\mu)_{3\times 5}&(+,+)\cdot (A_\mu)_{3\times 3}
    \end{array}
\right), 
 \\
  (\hat P_0,\hat P_1)\cdot (A_5)&=\left(
    \begin{array}{c|c}
(-,-)\cdot (A_5)_{5\times 5}&(+,+)\cdot (A_5)_{5\times 3}\\\cline{1-2}
(+,+)\cdot (A_5)_{3\times 5}&(-,-)\cdot (A_5)_{3\times 3}
    \end{array}
\right). 
\end{align}
The subscripts in the right-hand sides imply $n\times m$ submatrix in
the $SU(8)$ representation space. This boundary condition leads to
the residual symmetry $SU(5)\times SU(3)\times U(1)$.

Using the residual symmetry, we can simplify the VEVs of the zero mode
of $A_5$.  In this case the Wilson line phase degrees of freedom are
parametrized by the three real parameters $a_i$ ($i=1,2,3$) as
\begin{align}
  \vev{A_5}&=
{1\over 2gR}
\left(
    \begin{array}{c|c}
{\bf 0}_{5\times 5}&\Theta^b\\\cline{1-2}
(\Theta^b)^\dag&{\bf 0}_{3\times 3}
    \end{array}
\right), \qquad 
\Theta^b=
     \begin{pmatrix}
  a_1&0&0\\
  0&a_2&0\\
  0&0&a_3\\
  0&0&0\\
  0&0&0
     \end{pmatrix}.
\label{su8para}
\end{align}
If the parameters evolve non-zero VEVs of $a_1=a_2=a_3\neq 0$ (mod 1)
at a vacuum, then the spontaneous symmetry breaking
$SU(5)\times SU(3)\times U(1)\to G_{\rm SM}$ is realized.

From the parametrization in Eq.~\eqref{su8para}, we can see that
$a_1$, $a_2$, and $a_3$ correspond to generators involved in
$SU(2)_{16}$, $SU(2)_{27}$, and $SU(2)_{38}$, respectively.  In order
to derive the effective potential for the Wilson line phases, we
decompose $SU(8)$ representations into
$SU(2)_{16}\times SU(2)_{27}\times SU(2)_{38}$ representations.  The
$8$, $28$, $36$, and $63$-dimensional representations of $SU(8)$ are
decomposed as
\begin{align}
8&\ni 1\times [(2,1,1)+(1,2,1)+(1,1,2)],\\
28&\ni 2\times [(2,1,1)+(1,2,1)+(1,1,2)] 
  + 1\times [(2,2,1)+(1,2,2)+(2,1,2)],\\
\notag
36&\ni 
2\times [(2,1,1)+(1,2,1)+(1,1,2)] 
+ 1\times [(2,2,1)+(1,2,2)+(2,1,2)]\\
&\qquad +1\times [(3,1,1)+(1,3,1)+(1,1,3)] ,\\\notag
63&\ni 
4\times [(2,1,1)+(1,2,1)+(1,1,2)]+ 2\times [(2,2,1)+(1,2,2)+(2,1,2)] \\
&\qquad +1\times [(3,1,1)+(1,3,1)+(1,1,3)], 
\end{align}
where the right-hand sides indicate the
$(SU(2)_{16},SU(2)_{27},SU(2)_{38})$ irreducible representations.
From the expressions, as in the $SU(7)$ case, one can readily derive
contributions to the effective potential for $a_i$.

We denote the contributions to the effective potential from a bosonic
degree of freedom of the $R$-dimensional representation by
$F_8^{R}(a_i, \delta)$, which is written as follows:
\begin{align}
  F_8^{8}(a_i, \delta)&={-C\over 2}
\sum_{i=1}^3\csum(a_i+\delta),\\
  F_8^{28}(a_i, \delta)&={-C\over 2}
\left[
2\sum_{i=1}^3\csum(a_i+\delta)
+\sum_{1\leq i<j}^3\left\{\csum(a_i+a_j+\delta)+\csum(a_i-a_j+\delta)\right\}
\right],\\
  F_8^{36}(a_i, \delta)&=
{-C\over 2}
\left[
2\sum_{i=1}^3\csum(a_i+\delta)
+\sum_{1\leq i\leq j}^3
\left\{\csum(a_i+a_j+\delta)+\csum(a_i-a_j+\delta)\right\}
\right],\\
  F_8^{63}(a_i, \delta)&={-C\over 2}
\left[
4\sum_{i=1}^3\csum(a_i+\delta)
+\sum_{i,j=1}^3
\left\{\csum(a_i+a_j+\delta)+\csum(a_i-a_j+\delta)\right\}
\right].
\end{align}
We specify the numbers of the bulk fermion fields in this model by 
\begin{align}
{\mathcal N}_8\equiv (n_{8}^{(+)},n_{8}^{(-)},n_{28}^{(+)},n_{28}^{(-)},
n_{36}^{(+)},n_{36}^{(-)},n_{63}^{(+)},n_{63}^{(-)}), 
\end{align}
where we consider $R=8,28,36,63$ fermion fields having even $(+)$ 
or odd $(-)$ periodicities.
We obtain the one-loop effective potential in the $SU(8)$ model as 
\begin{align}\notag
  V_{8}(a_i, {\mathcal N}_8)
=&3F_8^{63}(a_i,0)
-4\bigg[
n_{8}^{(+)} F_8^{8}(a_i,0)
+n_{8}^{(-)} F_8^{8}(a_i,1)
+n_{28}^{(+)} F_8^{28}(a_i,0)
+n_{28}^{(-)} F_8^{28}(a_i,1)\\
&+n_{36}^{(+)} F_8^{36}(a_i,0)
+n_{36}^{(-)} F_8^{36}(a_i,1)
+n_{63}^{(+)} F_8^{63}(a_i,0)
+n_{63}^{(-)} F_8^{63}(a_i,1)\bigg].
\label{su8pot}
\end{align}

Vacuum configurations determined by the effective potentials in the
$SU(7)$ model in Eq.~\eqref{su7pot} and the $SU(8)$ model in
Eq.~\eqref{su8pot} depend on numbers of bulk fermion fields in each
model.  We will discuss the vacuum structures in the $SU(7)$ and
$SU(8)$ models in Sec.~\ref{vacan}.

\subsection{The $SO(10)$ model} 
\label{sec:so10}
In this subsection, we study another example of the gGHU model that
has the $SO(10)$ unified symmetry. In the $SO(10)$ model, the gauge
field $A_M$, which belongs to the 45-dimensional adjoint
representation, can be decomposed into the representations of the subgroup
$SU(5)$ as
\begin{align}
  A_M=A_M(24)+A_M(1)+A_M(10)+A_M(\overline{10}), 
\end{align}
where $A_M(R)$ transforms as the $R$-dimensional representation of $SU(5)$. 
The boundary condition is taken as follows:
\begin{align}
  \label{eq:8}
  (\hat P_0,\hat P_1)\cdot A_\mu(24)&={(+,+)}\cdot A_\mu(24),&
(\hat P_0,\hat P_1)\cdot A_\mu(1)&={(+,+)}\cdot A_\mu(1),\\
(\hat P_0,\hat P_1)\cdot A_\mu(10)&={(-,-)}\cdot A_\mu(10),&
(\hat P_0,\hat P_1)\cdot A_\mu(\overline{10})&={(-,-)}\cdot A_\mu(\overline{10}). 
\label{so10bc}
\end{align}
This leads to $SU(5)\times U(1)$ as the residual symmetry.

Since the extra-dimensional component of the gauge field has the
opposite parities to those of $A_\mu$ in Eqs.~\eqref{eq:8}
and~\eqref{so10bc}, there appears zero mode of $A_5$ in $A_5(10)$ and
$A_5(\overline{10})$.  Using the residual $SU(5)\times U(1)$ symmetry,
we can parametrize them by two real parameters $\til a$ and $\til b$
as
\begin{align}
\vev{A_5(10)}\equiv (H_{ij})&=
  \begin{pmatrix}
      0&H_{12}&H_{13}&H_{14}&H_{15}\\
      &0&H_{23}&H_{24}&H_{25}\\
      &&0&H_{34}&H_{35}\\
      &&&0&H_{45}\\
      &&&&0
  \end{pmatrix}
\to {1\over 2gR}
   \begin{pmatrix}
0&0&0&0&0\\       
&0&\til a&0&0\\       
&&0&0&0\\       
&&&0&\til b\\       
&&&&0
   \end{pmatrix}, 
\label{so10vev}
\end{align}
and $\vev{A_5(\overline{10})}=\vev{A_5({10})}^\dag$.  At a vacuum, if
one of the parameters takes a non-zero VEV (mod 1) and the other
remains zero (mod 2), then the spontaneous gauge symmetry breaking
$SU(5)\times U(1)\to G_{\rm SM}$ is realized.

We start to discuss the effective potential for $\til a$ and $\til b$.
In order to clarify the group structure, it is useful to consider the
decomposition $SU(5)\to SU(3)\times SU(2)\times U(1)'$, where
$\tilde a$ is a member of the triplet of the $SU(3)$ symmetry and
$\tilde b$ is a singlet under the $SU(3)\times SU(2)$ symmetry. Then
one can introduce a decomposition
$SO(10)\to SU(4)\times SU(2)\times SU(2)'$ where the above $SU(3)$ is
involved in $SU(4)$. In this basis, $A_5$ is written as
\begin{align}
  A_5=A_5((15,1,1))+A_5((1,3,1))+A_5((1,1,3))+A_5((6,2,2)), 
\label{422decomp}
\end{align}
where each term in the right-hand side transforms as
$(SU(4),SU(2),SU(2)')$ irreducible representations. Since $\til a$ is
involved in $\vev{A_5((15,1,1))}$, one can find an $SU(2)$ subgroup of
$SU(4)$ such that the parameter $\til a$ corresponds to a 
generator of the $SU(2)$ subgroup, which is referred to as $SU(2)_a$
in the following.  The other parameter $\til b$ is involved in
$\vev{A_5((1,1,3))}$.  It is clear that $\til b$ corresponds to
a generator of $SU(2)'$ of $SU(4)\times SU(2)\times SU(2)'$. Thus we
denote $SU(2)'=SU(2)_b$.

As in the previous subsection, we will derive the effective potential
for $\til a$ and $\til b$ focusing on $SU(2)_a\times SU(2)_b$ charges
of fields in the $SO(10)$ model. Under the 
$SO(10)\to SU(4)\times SU(2)\times SU(2)'$ decomposition, 
the $10$ and $16$-dimensional representations are written by 
\begin{align}
  10 &\to (6,1,1)+(1,2,2), \qquad   16 \to (4,2,1)+(\overline{4},1,2). 
\label{422decomp2}
\end{align}
From Eqs.~\eqref{422decomp} and~\eqref{422decomp2}, 
one can see that the $10$, $16$, and adjoint representations involve the following
$(SU(2)_a,SU(2)_b)$ representations:
\begin{align}
  10&\ni 2\times [(2,1) + (1,2)],\\
  16& \ni  2\times [(2,1) + (1,2)]+1\times (2,2),\\
  45&\ni   4\times [(2,1)+ (1,2)]+4\times (2,2)+1\times [(3,1)+ (1,3)].
\end{align}
The contributions to the effective potential for $\til a$ and $\til b$ from
a bosonic degree of freedom of the $R$-dimensional representation 
is denoted by $F_{10}^{R}(\til a, \til b, \delta)$; it is written as follows:
\begin{align}
\label{so10pot10}
  F_{10}^{10}(\til a, \til b, \delta)=&{-C\over 2}
\left[
2\csum (\til a+\delta)+2\csum (\til b+\delta)\right],\\
  F_{10}^{16}(\til a, \til b, \delta)=&{-C\over 2}
\left[2\csum (\til a+\delta)+2\csum (\til b+\delta)
\label{so10pot16}
+\csum (\til a+\til b+\delta)+\csum (\til a-\til b+\delta)
\right],\\
  F_{10}^{45}(\til a, \til b, \delta)=&{-C\over 2}
\left[4\csum (\til a+\delta)+4\csum (\til b+\delta)
+4\csum (\til a+\til b+\delta)
\label{so10pot45}
+4\csum (\til a-\til b+\delta)
  +\csum (2\til a+\delta)
+\csum (2\til b+\delta)
\right].
\end{align}

We use
${\mathcal N}_{10}=(n_{10}^{(+)},n_{10}^{(-)},
n_{16}^{(+)},n_{16}^{(-)},n_{45}^{(+)},n_{45}^{(-)})$
as the numbers of $R$-dimensional $(R=10,16,45)$ bulk fermion fields
of $\til\eta=\pm 1$ in this model.  Then the one-loop effective
potential has the following form:
\begin{align}\notag
  V_{10}(\til a,\til b, {\mathcal N}_{10})
&=3F_{10}^{45}(\til a,\til b,0)
-4\bigg[
n_{10}^{(+)} F_{10}^{10}(\til a,\til b,0)
+n_{10}^{(-)} F_{10}^{10}(\til a,\til b,1)\\
&\quad +n_{16}^{(+)} F_{10}^{16}(\til a,\til b,0)
+n_{16}^{(-)} F_{10}^{16}(\til a,\til b,1)
+n_{45}^{(+)} F_{10}^{45}(\til a,\til b,0)
+n_{45}^{(-)} F_{10}^{45}(\til a,\til b,1)\bigg].
\end{align}
We will discuss the vacuum structure of the potential in
Sec.~\ref{vacan}.

\bigskip
\section{The $E_6$ gGHU model} 
\label{sec:e6}
\subsection{Overview of the $E_6$ model} 
In this section, we study the gGHU model based on the $E_6$ unified
symmetry. This model leads to the gauge symmetry breaking
$E_6\to G_{\rm SM}$.  We give an overview of the model in this
subsection. The detailed structure of the model and the derivation of
the effective potential for the zero mode of $A_5$ is studied in the
following subsections.

We first summarize the group structure of $E_6$, which has three
maximal regular subgroups $SO(10)\times U(1)$, $SU(6)\times SU(2)$,
and $(SU(3))^3$~\cite{grouptheory}. We denote the subgroup $(SU(3))^3$
as $SU(3)_C\times SU(3)_L\times SU(3)_R$ where $SU(3)_C$ is identified
to the color gauge symmetry and $SU(3)_L$ involves the weak isospin
$SU(2)_L$.  An $SU(2)$ subgroup of $SU(3)_R$ is identified to
$SU(2)_R$ that is found in the Pati-Salam unified
symmetry~\cite{PS}. We take $SU(2)_R=SU(2)_{12}$, where $SU(2)_{ij}$
induces mixing between the $i$-th and $j$-th components of the
$SU(3)_R$ fundamental representation.  In addition, we refer to
$SU(2)_{23}$ and $SU(2)_{31}$ as $SU(2)_E$ and $SU(2)_{E_F}$,
respectively.  Among the $SU(2)$ symmetries in $SU(3)_R$, only
$SU(2)_E$ is orthogonal to $G_{\rm SM}$.

In the rest of the paper, we use the notation 
\begin{align}
  E_6
  ~\supset~ 
  \begin{cases}
  &SU(6)\times SU(2)_E
    ~\supset~ SU(5)\times U(1)_K\times SU(2)_E,\\
  &SU(6)_F\times SU(2)_{E_F}
    ~\supset~
    SU(5)_F\times U(1)_{K_F}\times SU(2)_{E_F}, 
  \end{cases}
    \label{su6su2def1}
\end{align}
where $SU(5)$ is the Georgi-Glashow unified symmetry that involves
$G_{\rm SM}$. Note that the maximal $SU(2)_R$ rotation, which we call
$SU(2)_R$ flip, corresponds to the exchange of the bases of
$SU(6)\times SU(2)_E$ and $SU(6)_F\times SU(2)_{E_F}$. 
Accordingly, $SU(5)_F$ is identified to the symmetry found 
in the flipped $SU(5)$ models~\cite{flipgut}.
We also use
\begin{align}
  E_6~\supset~ 
  &SO(10)\times U(1)_{V'}~\supset~
    \begin{cases}
        SU(5)\times U(1)_V\times U(1)_{V'},
\\
        SU(5)_F\times U(1)_{V_F}\times U(1)_{V'}. 
    \end{cases}
\label{su5u1u1def}
\end{align}
As in Eq.~\eqref{su6su2def1}, the $SU(2)_R$ flip leads to 
the exchange of the above two different bases
in the right-hand sides. 

In this model, the symmetry breaking is induced by a boundary
condition, the VEVs of Wilson line phases, and an anomaly.  As shown
below, the boundary condition leads to the symmetry breaking
$E_6\to SO(10)\times U(1)_{V'}$ at $y=0$ and
$E_6\to SU(6)_F\times SU(2)_{E_F}$ at $y=\pi R$. As a result, the
residual symmetry is $SU(5)_F\times U(1)_{V_F}\times U(1)_{V'}$.  The
zero mode of $A_5$ can develop VEVs, which lead to the symmetry
breaking $SU(5)_F\times U(1)_{V_F}\to G_{\rm SM}$.

We assume that the $U(1)_{V'}$ is broken by localized anomalies. The
orbifold allows us to obtain chiral fermions, and the fermion fields
generally contribute to anomalies at boundaries~\cite{5danom}.  In our
model, $U(1)_{V'}$ charges of the fermion fields that have the Neumann
boundary condition at $y=0$ tend to become anomalous. Localized
anomalies are assumed to be cancelled by the Green-Schwarz
mechanism~\cite{GS}.  Namely, a pseudo-scalar field that transforms
non-linearly under the $U(1)_{V'}$ symmetry and has the Wess-Zumino
couplings is introduced on this boundary to cancel the anomaly. The
scalar field allows a mass term~\cite{Stuckelberg} for the $U(1)_{V'}$
gauge field on the boundary.  Thus we also assume that there appears a
localized heavy mass term for the $U(1)_{V'}$ gauge field at the
boundary $y=0$ due to the $U(1)_{V'}$ breaking.

In the following subsections, we will show the explicit formulations
of the $E_6$ model. Then we will derive the contributions to the
effective potential for $A_5$ from bulk fermion fields and the gauge
field taking into account the effect of the localized mass term on
the effective potential.

\subsection{The boundary condition and the $A_5$ zero mode} 
\label{subsec:e6bcs}
In the $E_6$ model, in order to show the boundary condition, we
decompose the gauge field $A_M$, which belongs to the 78-dimensional
adjoint representation of $E_6$, into $SO(10)$ and
$SU(6)_F\times SU(2)_{E_F}$ representations as
\begin{align}
\label{e6am1}
    A_M&=A_M(45)+A_M(1)+A_M(16)+A_M(\overline{16}) \\ 
  &=A_M((35,1))+A_M((1,3))+A_M((20,2)), 
\label{e6am2}
\end{align}
where each term in the right-hand sides transforms as the $SO(10)$ or
$SU(6)_F\times SU(2)_{E_F}$ irreducible representations.

We introduce the following boundary condition: 
\begin{align}
\label{so10dec}
    \hat P_0\cdot A_\mu&=+ A_\mu(45)+A_\mu(1)-A_\mu(16)-A_\mu(\overline{16}), \\
\label{su6dec}
  \hat P_1\cdot A_\mu&=+ A_\mu((35,1))+A_\mu((1,3))-A_\mu((20,2)). 
\end{align}
The residual symmetry is $SU(5)_F\times U(1)_{V_F}\times U(1)_{V'}$.

It is useful to decompose the fields in Eq.~\eqref{e6am1} into
$SU(5)_F\times U(1)_{V_F}$ representations as 
\begin{align}
  \label{su5u1decomp1}
 A_\mu(45)&=A_\mu(24_0)^{(+,+)}+A_\mu(1^{V_F}_0)^{(+,+)}
            +A_\mu(10_{-4})^{(+,-)}+A_\mu(\overline{10}_{4})^{(+,-)},\\
  A_\mu(1)&=A_\mu(1^{V'}_0)^{(+,+)},\\
  A_\mu(16)&=A_\mu(10_1)^{(-,-)}+A_\mu(\overline{5}_{-3})^{(-,+)}
             +A_\mu(1_5)^{(-,+)},\\
  A_\mu(\overline{16})&=A_\mu(\overline{10}_{-1})^{(-,-)}
                        +A_\mu(5_{3})^{(-,+)}+A_\mu(1_{-5})^{(-,+)},
  \label{su5u1decomp4}
\end{align}
where $A_\mu (R_Q)$ in the right-hand sides corresponds to the field
that transforms as the $R$-dimensional $SU(5)_F$ representation and
has the $U(1)_{V_F}$ charge $Q$. The superscript indicates orbifold
parity $(\hat P_0,\hat P_1)$ of each field.  Note that due to the
localized mass term for the $U(1)_{V'}$ gauge field at the boundary
$y=0$, the orbifold parities of the gauge field are effectively
modified. As a result, they generally obey a mixed boundary
condition~\cite{mixedbc}; the effect of the modification will be
discussed in Sec.~\ref{subsec:effanom}. The above
$SU(5)_F\times U(1)_{V_F}$ representations are related to
$SU(6)_F\times SU(2)_{E_F}$ representations in Eq.~\eqref{e6am2} as
follows:
\begin{align}
  \label{su5decomp21}
 A_\mu((35,1))&=A_\mu(24_0)^{(+,+)}+A_\mu(5_{3})^{(-,+)}
                +A_\mu(\overline{5}_{-3})^{(-,+)}
                +A_\mu(1^{K_F}_0)^{(+,+)},\\
  A_\mu((1,3))&=A_\mu(1_5)^{(-,+)}+A_\mu(1_{-5})^{(-,+)}+A_\mu(1^{E_F}_0)^{(+,+)},\\
  A_\mu((20,2))&=A_\mu(10_1)^{(-,-)}+A_\mu(\overline{10}_{-1})^{(-,-)}
                 +A_\mu(10_{-4})^{(+,-)}+A_\mu(\overline{10}_{4})^{(+,-)}, 
  \label{su5decomp23}
\end{align}
where $A_\mu(1^{K_F}_0)^{(+,+)}$ and $A_\mu(1^{E_F}_0)^{(+,+)}$ are
linear combinations of $A_\mu(1^{V_F}_0)^{(+,+)}$ and
$A_\mu(1^{V'}_0)^{(+,+)}$. 

Note that $A_5(R_Q)$ has opposite orbifold parities to those of
$A_\mu(R_Q)$. Thus the zero mode of $A_5$ appears in
$A_5(10_1)^{(+,+)}$ and $A_5(\overline{10}_{-1})^{(+,+)}$. With the
help of the residual $SU(5)_F\times U(1)_{V_F}$ symmetry, the VEVs of
the zero mode are simplified as follows:
\begin{align}
\vev{A_5(10_1)^{(+,+)}}
\to {1\over 2gR}
   \begin{pmatrix}
0&0&0&0&0\\       
&0&\til d&0&0\\       
&&0&0&0\\       
&&&0&\til n\\       
&&&&0
   \end{pmatrix}, 
\label{wdofe6}
\end{align}
where $\til d$ and $\til n$ are real parameters, and
$\vev{A_5(\overline{10}_{-1})^{(+,+)}}=\vev{A_5(10_1)^{(+,+)}}^\dag$.
The parameter $\til d$ ($\til n$) corresponds to a generator of an
$SU(2)$ subgroup of $E_6$, which is referred to as $SU(2)_d$
($SU(2)_n$) in the following.  If one of the parameters takes a
non-trivial VEV (mod 1) and the other remains zero (mod 2), then the
symmetry breaking $SU(5)_F\times U(1)_{V_F}\to G_{\rm SM}$ is realized
similarly to the previous $SO(10)$ model.

\subsection{Contributions to the effective potential from bulk fermion fields} 
\label{subsec:a5eff}
We start to derive the effective potential for the zero mode of $A_5$,
namely parameters $\til d$ and $\til n$ in Eq.~\eqref{wdofe6}. The
effective potential is generated by quantum corrections from matter
and the gauge fields in the model.  We here focus on the contributions to
the effective potential from bulk fermion fields; the contributions
from the gauge field are studied in the next subsection.

The contributions in the $E_6$ model can be easily obtained from the
result in Sec.~\ref{sec:so10}. To see this, it is required to find
another $SO(10)$ subgroup of $E_6$ where the parameters $\til d$ and
$\til n$ in Eq.~\eqref{wdofe6} belongs to the $45$-dimensional adjoint
representation. For this purpose, we consider maximal $SU(2)_{E_F}$
rotation of the $SO(10)\times U(1)_{V'}$ decompositions in
Eqs.~\eqref{su5u1decomp1}--\eqref{su5u1decomp4}. This leads to a new
basis $E_6\supset SO(10)'\times U(1)_{V''}$, where
Eqs.~\eqref{su5u1decomp1}--\eqref{su5u1decomp4} are changed to
\begin{align}
\label{so10p45}
  A_\mu(45')&=A_\mu(24_0)^{(+,+)}+A_\mu(1^{V_F'}_0)^{(+,+)}+A_\mu(10_1)^{(-,-)}
              +A_\mu(\overline{10}_{-1})^{(-,-)},\\
  A_\mu(1')&=A_\mu(1^{V''}_0)^{(+,+)},\\
  A_\mu(16')&=A_\mu(10_{-4})^{(+,-)}+A_\mu(\overline{5}_{-3})^{(-,+)}
              +A_\mu(1_5)^{(-,+)},\\
  A_\mu(\overline{16}')&=A_\mu(\overline{10}_{4})^{(+,-)}
                         +A_\mu(5_{3})^{(-,+)}+A_\mu(1_{-5})^{(-,+)}.
\label{so10p452}
\end{align}
In this expression, the left-hand sides transform as the irreducible
representations of $SO(10)'$, and $A_\mu(1^{V_F'}_0)^{(+,+)}$ and
$A_\mu(1^{V''}_0)^{(+,+)}$ are linear combinations of
$A_\mu(1^{V_F}_0)^{(+,+)}$ and $A_\mu(1^{V'}_0)^{(+,+)}$. Note that
the fields in $A_\mu(45')$ have the same orbifold parities as those in
Eqs.~\eqref{eq:8}~and~\eqref{so10bc}. This coincides with the fact
that the zero mode of $A_5$ parametrized by $\til d$ and $\til n$
appears in the adjoint representation $A_5(45')$ of $SO(10)'$. In
addition, since the gauge interactions respect the $SO(10)'$ symmetry
and the orbifold parities, other $SO(10)'$ representations in the
$E_6$ model should have the same orbifold parities as in the previous
$SO(10)$ model in Sec.~\ref{sec:so10} up to overall signs. These
observations imply that the contributions to the effective potential
in the $E_6$ model can be written by using the contributions in the
$SO(10)$ model.

Let us consider the contribution from a bulk adjoint fermion
$\Phi_A^{(\til \eta)}$ having $\til\eta=\eta_0\eta_1$, which is
decomposed into the $SO(10)'$ multiplets as
$\Phi_A^{(\til \eta)} = \Phi_A^{(\til \eta)}(45')+\Phi_A^{(\til
  \eta)}(1')+\Phi_A^{(\til \eta)}(16') +\Phi_A^{(\til
  \eta)}(\overline{16}')$.  One can see that, for instance, the
orbifold parity of $\Phi_A^{(+1)}(45')$ $(\Phi_A^{(+1)}(16'))$
corresponds to one of the adjoint field with $\til \eta=+1$ (16-plet
with $\til \eta=-1$) in the previous $SO(10)$ model. As in the
previous section, we denote the contribution from a field that has a
bosonic degree of freedom of the $R$-dimensional representation
($R=27, 78$) by $F^{R}(\til d,\til n,\delta)$. From the above
discussion, the contribution $F^{78}(\til d,\til n,\delta)$ is shown
by the terms in Eqs.~\eqref{so10pot10}--\eqref{so10pot45} as
\begin{align}
  F^{78}(\til d,\til n,0)&=F_{10}^{45}(\til d, \til n, 0)+2F_{10}^{16}(\til d, \til n, 1),\\
  F^{78}(\til d,\til n,1)&=F_{10}^{45}(\til d, \til n, 1)+2F_{10}^{16}(\til d, \til n, 0). 
\end{align}
The explicit form of the contribution is 
\begin{align}
 \notag
  F^{78}(\til d, \til n, \delta)=&{-C\over 2}
\bigg[
4\csum(\til d)+4\csum(\til d+1)
+4\csum(\til n) +4\csum(\til n+1) 
+2\csum(\til d+\til n)
+2\csum(\til d+\til n+1)
+2\csum(\til d-\til n)\\ &
+2\csum(\til d-\til n+1)
+2\csum(\til d+\til n+\delta)
+2\csum(\til d-\til n+\delta)
+\csum(2\til d+\delta)
+\csum(2\til n+\delta)
\bigg].
\label{potf78n}  
\end{align}

Similar discussion holds for the contributions from a bulk 27-plet
fermion $\Phi_F^{(\til \eta)}$ having $\til\eta=\eta_0\eta_1$.  In
this model, the orbifold parities of the 27-plet are
\begin{align}
  \hat P_0\cdot \Phi_F^{(\til \eta)}&=
\eta_0\cdot \left\{-\Phi_F^{(\til \eta)}(16)
+\Phi_F^{(\til \eta)}(10)
+\Phi_F^{(\til \eta)}(1)\right\}, \\
  \hat P_1\cdot \Phi_F^{(\til \eta)}&=
\eta_1\cdot \left\{+\Phi_F^{(\til \eta)}((15,1))
-\Phi_F^{(\til \eta)}((\overline{6},2))\right\}, 
\end{align}
where each of the terms in the right-hand sides is the irreducible
representation of $SO(10)\times U(1)_{V'}$ or
$SU(6)_F\times SU(2)_{E_F}$. One can find the
$SO(10)\to SU(5)_F\times U(1)_{V_F}$ decomposition:
\begin{align}
\Phi_{F}^{(\til \eta)}(16)&=
\Phi_{F}(10_1)^{(-\eta_0,+\eta_1)}+\Phi_{F}(\overline{5}_{-3})^{(-\eta_0,-\eta_1)}+
\Phi_{F}(1_5)^{(-\eta_0,-\eta_1)}, \\
\Phi_{F}^{(\til \eta)}(10)&=
\Phi_{F}(5_{-2})^{(+\eta_0,+\eta_1)}+\Phi_{F}(\overline{5}_{2})^{(+\eta_0,-\eta_1)}, \\
\Phi_{F}^{(\til \eta)}(1)&=
\Phi_{F}(1_0)^{(+\eta_0,-\eta_1)}, 
\end{align}
where $\Phi_{F}(R_Q)$ in the right-hand sides means the field
transforms as the $R$-dimensional representation of $SU(5)_F$ and has
the $U(1)_{V_F}$ charge $Q$. The superscript in the right-hand sides
indicates the eigenvalues of the parity $(\hat P_0,\hat P_1)$ of each
field. Using the $SU(2)_{E_F}$ rotation, one can obtain the
$SO(10)'\times U(1)_{V''}$ decomposition:
\begin{align}
\Phi_{F}^{(\til \eta)}(16')&=
\Phi_{F}(10_1)^{(-\eta_0,+\eta_1)}+\Phi_{F}(\overline{5}_{2})^{(+\eta_0,-\eta_1)}+
\Phi_{F}(1_0)^{(+\eta_0,-\eta_1)}, \\
\Phi_{F}^{(\til \eta)}(10')&=
\Phi_{F}(5_{-2})^{(+\eta_0,+\eta_1)}+\Phi_{F}(\overline{5}_{-3})^{(-\eta_0,-\eta_1)}, \\
\Phi_{F}^{(\til \eta)}(1')&=
\Phi_{F}(1_5)^{(-\eta_0,-\eta_1)}. 
\end{align}
From the expressions, one can obtain the contributions to the
effective potential from the 27-plet as
\begin{align}
  F^{27}(\til d,\til n,0)&=F_{10}^{16}(\til d, \til n, 1)+F_{10}^{10}(\til d, \til n, 0),\\
  F^{27}(\til d,\til n,1)&=F_{10}^{16}(\til d, \til n, 0)+F_{10}^{10}(\til d, \til n, 1),
\end{align}
where terms in the right-hand sides are found in
Eqs.~\eqref{so10pot10}--\eqref{so10pot45}.  More explicitly, we obtain
\begin{align}
  F^{27}(\til d, \til n, \delta)=&{-C\over 2}
\bigg[
2\csum(\til d)
+2\csum(\til d+1)
+2\csum(\til n)
+2\csum(\til n+1)
+\csum(\til d+\til n+1-\delta)
+\csum(\til d-\til n+1-\delta)
\bigg].
\label{potf27p}
\end{align}
In Appendix~\ref{sec:apml}, we also derive the effective potential and
confirm the result in Eqs.~\eqref{potf78n} and~\eqref{potf27p} by
using another explicit formulation.

Several comments are in order.  In our setup, $\til d$ and $\til n$
parametrize the Wilson line phase degrees of freedom. Thus the
contributions to the effective potential in Eqs.~\eqref{potf78n}
and~\eqref{potf27p} are invariant under discrete shift of the
parameters as $\til n\to \til n+2$ or $\til d\to \til d+2$. In
addition, the residual $SU(5)$ symmetry of the model ensures that the
contributions are invariant under the changes of signs of $\til n$ and
$\til d$ and the exchange of values of $\til n$ and $\til d$.  Similar
invariance is found in the $SU(7)$, $SU(8)$, and $SO(10)$ models. In
addition, in this $E_6$ model, the contributions are also unchanged
under the transformation $(\til n,\til d)\to (\til n+1,\til d+1)$.
This invariance is not accidental at the one-loop level but guaranteed
by the $E_6$ symmetry in this model as shown in
Appendix~\ref{sec:mod}.

\subsection{Contributions to the effective potential from the gauge field} 
\label{subsec:effanom}
In this subsection, we will discuss the contributions to the effective
potential from the gauge field, whose $U(1)_{V'}$ component has a
localized mass term at the $y=0$ boundary due to the anomaly
cancellation. In order to do this, we need to show the KK mass
spectrum, which depends on the boundary mass parameter and the
background fields $(\til d,\til n)$, of the gauge field.  The detailed
derivation of the mass spectrum is shown in
Appendix~\ref{sec:apanom}. We here shortly summarize the calculation
procedure and the result of the calculation of the effective
potential.

The KK mass spectrum is obtained with the solution of equations of
motion (EOM) of the gauge field in the bulk. The bulk EOM is
simplified in the basis of $SU(6)\times SU(2)_E$, where the
$U(1)_{V'}$ component $A_\mu(V')$ of the gauge field is written by a
linear combination of the components $A_\mu(n^{(3)})$,
$A_\mu(d^{(3)})$, and $A_\mu(X)$, where $n^{(3)}$ $(d^{(3)})$ implies
the Cartan generator of $SU(2)_{n}$ ($SU(2)_{d}$), which is defined
below Eq.~\eqref{wdofe6}, and $X$ is a generator involved in
$SU(6)/SU(2)_d$. Since the parameter $\til n$ $(\til d)$ appears in
the bulk EOM, $A_\mu(n^{(3)})$ $(A_\mu(d^{(3)}))$ mixes with another
$SU(2)_{n}$ $(SU(2)_{d})$ gauge field component, which we call
$A_\mu(n^{(2)})$ $(A_\mu(d^{(2)}))$. Therefore, as discussed in
sec.~\ref{sec:apanom}, we should solve the EOM as simultaneous
equations concerning the set of fields
$(A_\mu(n^{(3)}), A_\mu(n^{(2)}), A_\mu(d^{(3)}), A_\mu(d^{(2)}),
A_\mu(X))$.

To obtain the solution of the EOM, the boundary condition should be
imposed. The boundary condition at the fixed points $y=0$ and $\pi R$
corresponds to $(\hat P_0,\hat P_1)$ parities as in
Eqs.~\eqref{su5u1decomp1}--\eqref{su5u1decomp4}, where even (odd)
parity means the Neumann (Dirichlet) boundary condition. In addition
the boundary condition for $A_\mu(V')$ is effectively modified from
$(\hat P_0,\hat P_1)$, since it has a localized mass term at the
boundary $y=0$~\cite{mixedbc}. Note that if the localized mass scale
is much larger than the compactification scale $1/R$, the Neumann
boundary condition of $A_\mu(V')$ at $y=0$ is effectively modified to
the Dirichlet boundary condition. With this heavy mass limit, the mass
spectrum of the $n$-th KK mode of the fields
$(A_\mu(n^{(3)}), A_\mu(n^{(2)}), A_\mu(d^{(3)}), A_\mu(d^{(2)}),
A_\mu(X))$ has the following form:
\begin{align}\label{mkk1}
    m_{\rm KK}^{(n)2}=\left({n+\rho_+\over R}\right)^2,
\qquad  &\rho_+={1\over \pi}\arcsin\sqrt{S_1+S_2\over 8},
\qquad n=0,\pm 1,\pm 2,\cdots,\\\label{mkk2}
  m_{\rm KK}^{(n)2}=\left({n+\rho_-\over R}\right)^2,
\qquad &\rho_-={1\over \pi}\arcsin\sqrt{S_1-S_2\over 8},
\qquad n=0,\pm 1,\pm 2,\cdots, \\\label{s1s2}
S_1={5\over 2}\left[\sin^2(\pi\til d )+\sin^2(\pi\til n )\right],\quad 
S_2&={5\over 2}\left[\left(\sin^2(\pi\til d )-\sin^2(\pi\til n )\right)^2
+{36\over 25}\sin^2(\pi\til d )\sin^2(\pi\til n )\right]^{1/2}, 
\end{align}
in addition to a $\til d$, $\til n$-independent mass.

For the extra-dimensional component $A_5$, there is no direct coupling
to the localized mass parameter. However, the boundary condition of $A_5$
could be modified in accordance with the modification of the boundary
condition of $A_\mu$ due to a gauge fixing term, which mixes $A_5$
with $A_\mu$. We demonstrate how the boundary condition of $A_5$ is
modified in a simple setup in Appendix~\ref{sec:bcmod}.

With the KK mass spectrum of the gauge fields in Eqs.~\eqref{mkk1}
and~\eqref{mkk2}, we can easily derive the contributions to the effective
potential from the gauge sector in this model. The contribution from a
bosonic degree of freedom is
\begin{align}\notag
F^{\rm A}(\til d,\til n)=&{-C\over 2}
\bigg[
4\csum(\til d)+4\csum(\til d+1)
+4\csum(\til n)+4\csum(\til n+1)
+4\csum(\til d+\til n)+4\csum(\til d-\til n)\\&
+2\csum(\til d+\til n+1)+2\csum(\til d-\til n+1)
+\csum(2\rho_+)+\csum(2\rho_-)
\bigg].  
\label{fa}
\end{align}

Using the contributions in Eqs.~\eqref{potf78n}, \eqref{potf27p},
and~\eqref{fa}, we can express the effective potential in the $E_6$
model. The matter content of the model is specified by $n_R^{(\pm)}$
$(R=27,78)$ that is a number of $R$-dimensional bulk fermion fields
with $\til\eta=\pm 1$, and we use
${\mathcal N}_{E_6}=(n_{27}^{(+)},n_{27}^{(-)},
n_{78}^{(+)},n_{78}^{(-)})$.
The one-loop effective potential for $\til d$ and $\til n$ is
\begin{align}
  V_{E_6}(\til d,\til n, {\mathcal N}_{E_6})
&=3F^{\rm A}(\til d,\til n)
-4\bigg[
n_{27}^{(+)} F^{27}(\til d,\til n,0)
+n_{27}^{(-)} F^{27}(\til d,\til n,1) 
+n_{78}^{(+)} F^{78}(\til d,\til n,0)
+n_{78}^{(-)} F^{78}(\til d,\til n,1)
\bigg]. 
\end{align}

\bigskip
\section{Analysis of vacuum structure} 
\label{vacan}
In this section, we study vacuum structure of the one-loop effective
potentials in the $SU(7)$, $SU(8)$, $SO(10)$, and $E_6$ models
discussed in the previous sections. In general, positions of the vacua
of the potentials depend on bulk matter contents of the
models. Without bulk matter fields, for instance, Wilson line phase
degrees of freedom do not evolve VEVs at the vacua in each model. In
this case, the residual gauge symmetries survive at the energy scale
well below $1/R$. On the other hand, if there are bulk fields
that lead to non-zero VEVs of the phases, then the residual
symmetries are further broken around the compactification scale. In
view of the gGHU scenario, we are interested in the case where
low-energy symmetries are compatible with the standard model gauge
group.

In each model, we found matter contents for bulk fermion fields
that lead to the symmetry breaking into $G_{\rm SM}$. Examples are
summarized in Table~\ref{tab:vac}, where the matter contents and the
VEVs at vacua are shown. In addition, we show the physical mass
spectrum of the zero mode of $A_5$ in a normalization, whose
definition is expressed in Eq.~\eqref{su7massmat}. In all the models,
bulk fermions that have periodic boundary conditions ($\til \eta=0$)
are required to obtain non-zero VEVs.
\begin{table}
\begin{center}
\caption{Examples of vacuum configurations in $SU(7)$, $SU(8)$, $SO(10)$, and $E_6$ models. 
Matter contents, VEVs at the global minimum of the one-loop effective potentials, and 
physical squared mass eigenvalues of $A_5$ zero modes normalized by the typical mass parameter $m_0^2$ 
in Eq.~\eqref{su7massmat} are shown.
}
\bigskip
\label{tab:vac}
\begin{tabular}{clll} 
\hline\hline
    model &\multicolumn{1}{c}{matter contents}&\multicolumn{1}{c}{VEVs} & 
\multicolumn{1}{c}{normalized squared masses}\\\hline
    $SU(7)$ &${\mathcal N}_7=(0,0,0,0,2,0,0,0)$ & $a_1=a_2=0.5849$ &$(105.4,~65.76)$ \\
    $SU(7)$ &${\mathcal N}_7=(2,0,0,2,0,0,2,0)$ & $a_1=a_2=0.6667$ &$(126.5,~58.00)$ \\
    $SU(8)$ &${\mathcal N}_8=(0,0,0,0,2,0,0,0)$ & $a_1=a_2=a_3=0.5490$ &$(125.0,~63.64,~63.64)$ \\
    $SU(8)$ &${\mathcal N}_8=(2,0,0,2,0,0,2,0)$ & $a_1=a_2=a_3=0.6482$ &$(139.8,~39.79,~39.79)$ \\
    $SO(10)$ &${\mathcal N}_{10}=(0,1,0,0,1,0)$ & $\til a=0$,~ $\til b=0.7205$ &$(20.82,~13.43)$ \\
    $SO(10)$ &${\mathcal N}_{10}=(0,2,0,0,1,0)$ & $\til a=0$,~ $\til b=0.3103$ &$(26.36,~5.008)$ \\
    $E_6$ &${\mathcal N}_{E_6}=(0,2,1,2)$ & $\til d=0$,~ $\til n=0.0825$ &$(21.33,~3.260)$ \\
    $E_6$ &${\mathcal N}_{E_6}=(0,4,1,2)$ & $\til d=0$,~ $\til n=0.6874$ &$(61.20,~36.91)$ \\\hline\hline
  \end{tabular} 
\end{center}
\end{table}

\begin{figure}[t]
\begin{center}
  \includegraphics[width=4.5cm,clip]{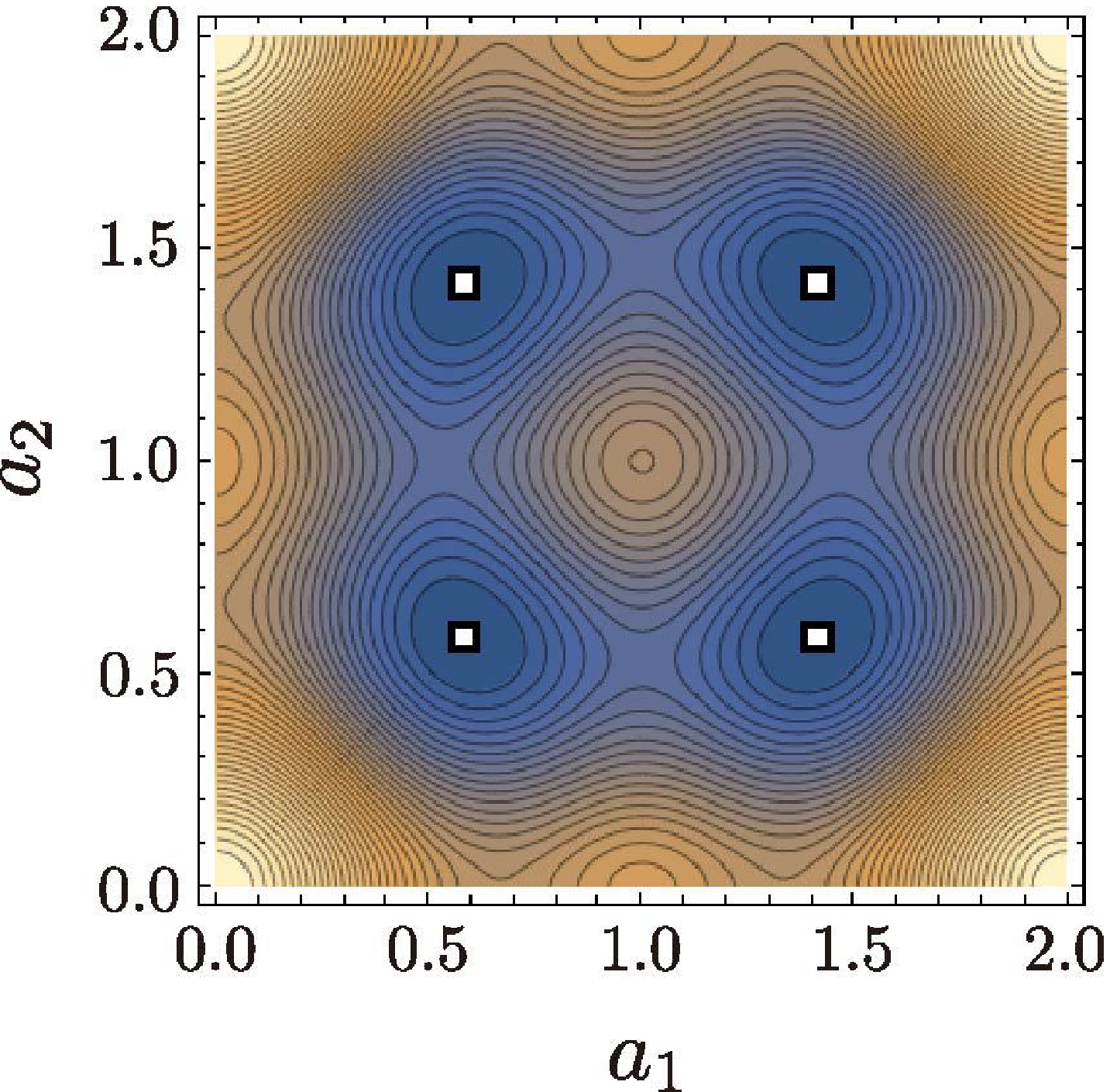}\hspace*{0.7cm}
  \includegraphics[width=4.5cm,clip]{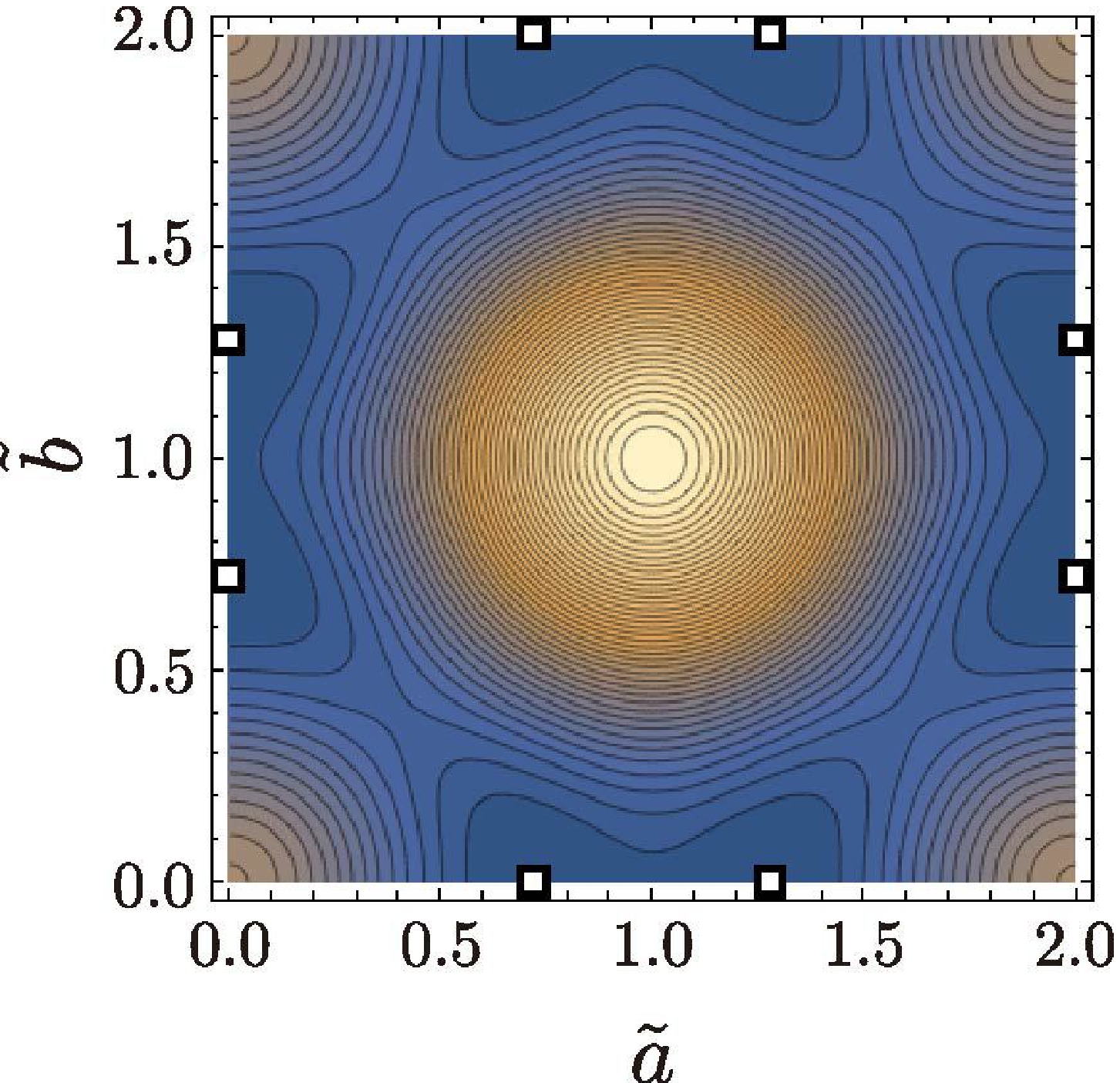}\hspace*{0.7cm}
  \includegraphics[width=4.5cm,clip]{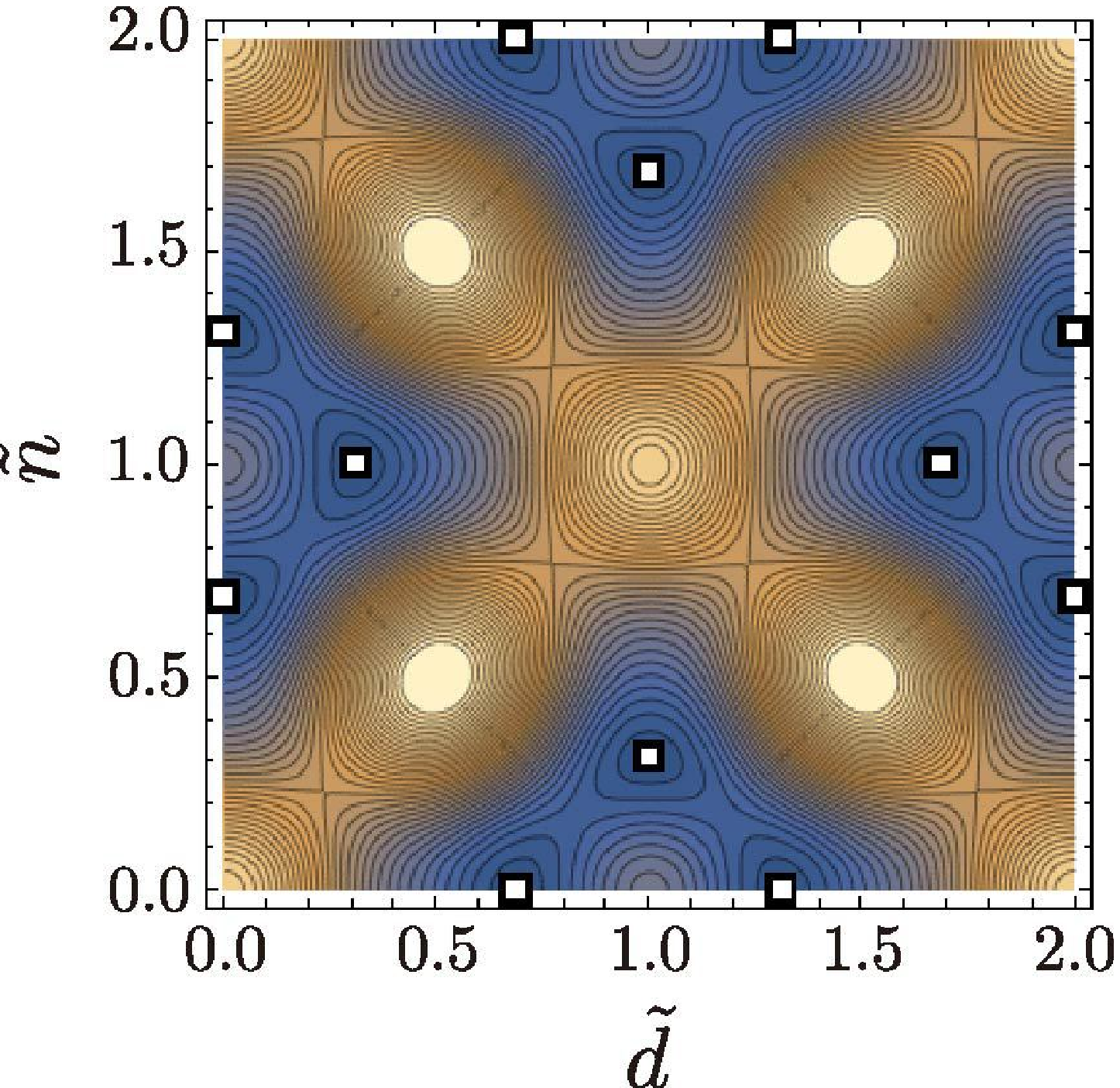}
\caption{
The color contour plots of the one-loop effective potentials as functions of the 
parametrized zero modes of $A_5$ in $SU(7)$ model with 
${\mathcal N}_7=(0,0,0,0,2,0,0,0)$ (left), 
$SO(10)$ model with ${\mathcal N}_{10}=(0,1,0,0,1,0)$ (center), and 
$E_6$ model with ${\mathcal N}_{E_6}=(0,4,1,2)$ (right). 
From light orange region to dark blue region, values of the potential decrease. 
In each figure, we show positions of the global minima of the potential by the square symbols. 
}
\label{fig:su7so10e6} 
\end{center}
\end{figure}

In the $SU(7)$ model, non-zero VEVs $a_1=a_2\neq 0$ $({\rm mod}~1)$
are obtained at vacua, where the residual symmetry
$SU(5)\times SU(2)\times U(1)$ is broken to $G_{\rm SM}$.
Figure~\ref{fig:su7so10e6} left shows the contour plot of the one-loop
effective potential $V_{7}(a_i, {\mathcal N}_7)$ for the case with
${\mathcal N}_7=(0,0,0,0,2,0,0,0)$. The positions of the vacua are
denoted by the square symbols in the contour plot. Under the shift
$a_i\to a_i+2$ for each $i$, the potential has invariance, which
reflects the phase property of each $a_i$. Also one can see the
potential is invariant under $a_i\to -a_i$ or
$(a_1,a_2)\to (a_2,a_1)$. In this figure, one can see that there
appear four degenerate vacua, which are physically equivalent. Around
the vacua, the physical zero mode of $A_5$ becomes massive and the
mass matrix is evaluated as 
\begin{align}
  (M_7^2)_{jk}&=
m_0^2 {\del^2 \over \del a_k \del a_j}{V_7(a_i,{\mathcal N}_7)\over C}, 
\qquad m_0^2={3g_{4D}^2\over 32\pi^6 R^2}, 
\qquad g_{4D}={g\over \sqrt{2\pi R}}, 
\label{su7massmat}
\end{align}
where we introduce $m_0^2$ as a typical (squared) mass scale and the
effective four-dimensional gauge coupling $g_{4D}$. In the present
case, at one of the vacua, the VEVs take $a_1=a_2=0.5849$ and the
eigenvalues of the mass matrix are $105.4 m_0^2$ and $65.76 m_0^2$.
The values in Table~\ref{tab:vac} show the eigenvalues of squared
masses normalized by $m_0^2$.  For the case with
${\mathcal N}_7=(2,0,0,2,0,0,2,0)$, we also show the VEVs and squared
mass eigenvalues normalized by $m_0^2$ in Table~\ref{tab:vac}.

In the $SU(8)$ model, the residual symmetry is
$SU(5)\times SU(3)\times U(1)$, which is broken to $G_{\rm SM}$ by
VEVs $a_1=a_2=a_3\neq 0$ $({\rm mod}~1)$. For instance, the symmetry
breaking is achieved for the cases with
${\mathcal N}_8=(0,0,0,0,2,0,0,0)$ and
${\mathcal N}_8=(2,0,0,2,0,0,2,0)$.  Around the vacua, the mass matrix
of the physical mass spectrum of the zero mode of $A_5$ is evaluated
from $V_8(a_i,{\mathcal N}_8)$ similarly to Eq.~\eqref{su7massmat}.
In Table~\ref{tab:vac}, the VEVs and squared
mass eigenvalues are shown for the above two cases. In the
eigenvalues, there appears degeneracy, which reflects that two linear
combinations of the parametrized VEVs $a_{1,2,3}$ belong to the adjoint
representation of $SU(3)_C$ in $G_{\rm SM}$.

The $SO(10)$ model has the residual symmetry $SU(5)\times U(1)$.  For
the cases where one of the parameters $\til a$ and $\til b$ in
Eq.~\eqref{so10vev} has a non-zero VEV (mod 1) while the other remains
zero (mod 2), then $G_{\rm SM}$ is obtained. In Table~\ref{tab:vac},
we show the examples of the matter contents and corresponding VEVs
that lead to $G_{\rm SM}$. For the case with
${\mathcal N}_{10}=(0,1,0,0,1,0)$, we also show the contour plot of
the effective potential in Figure~\ref{fig:su7so10e6} center, where
the square symbols indicate the positions of degenerate vacua, which are
physically equivalent. Around the vacua, one can evaluate the physical
mass spectrum of $A_5$ using the potential
$V_{10}(\til a,\til b, {\mathcal N}_{10})$ similarly to
Eq.~\eqref{su7massmat}. The eigenvalues of the squared masses are also
shown in Table~\ref{tab:vac}.

Finally we discuss the $E_6$ model, where $\til d$ and
$\til n$ in Eq.~\eqref{wdofe6} parametrize the VEVs. The residual
symmetry of the model is $SU(5)_F\times U(1)_{V_F}$.  We are
particularly interested in the vacua that lead to $G_{\rm SM}$, where
VEVs are $\til d=0$ (mod 2) and $\til n\neq 0$ (mod 1).  We show two cases
with ${\mathcal N}_{E_6}=(0,2,1,2)$ and ${\mathcal N}_{E_6}=(0,4,1,2)$
in Table~\ref{tab:vac}. For the latter case, the contour plot of
$V_{E_6}(\til d,\til n, {\mathcal N}_{E_6})$ is also shown in
Figure~\ref{fig:su7so10e6} right. One can confirm that the potential
has the invariance that was mentioned in Sec.~\ref{subsec:a5eff}.
Around the vacua of the potential, physical components of the zero
mode of $A_5$ become massive.  The VEVs and the squared mass eigenvalues
are shown in Table~\ref{tab:vac}.

\bigskip
\section{Summary and discussion} 
We have studied the five-dimensional gGHU models, namely the
applications of the Hosotani mechanism to the unified gauge symmetry
breaking, on the orbifold compactification $S^1/\mathbb{Z}_2$. In
these models, VEVs of the zero modes of the extra-dimensional gauge
fields, whose dynamics reflects degrees of freedom of Wilson line
phases, are available to break the residual symmetry. The effective
potential of the zero mode is generated by quantum corrections that
depend on matter contents in the models.

We have discussed the models based on $SU(7)$, $SU(8)$, $SO(10)$, and
$E_6$ gauge symmetries. In each model, the standard model gauge
symmetry is achieved at a low-energy regime when the suitable bulk
fermion fields are contained. We have derived the one-loop effective
potentials for the zero modes of the extra-dimensional gauge fields in
all the models. We have also studied the effects of the localized mass
term for the gauge field induced by the anomaly in the $E_6$ model.

We have analyzed vacuum structures of the effective potentials and
shown examples of the matter contents for bulk fermion fields that
lead to the non-trivial VEVs of the zero modes and the standard model
gauge symmetry at the vacua. Our discussions have shown that not
so many bulk fermions are required to achieve the desired vacua; it
implies that the unified symmetry is naturally broken by dynamics of
the Wilson line phases, owing to extra dimensions in our Universe. 

To make our discussion more concrete, the mass spectrum of the bulk
fermion fields and the standard model matter sector should be
explicitly treated. In this case, one can examine the renormalization
group evolution of the gauge coupling constants, which has large
dependence on the bulk fermion mass spectrum. Since the mass spectrum
is not severely constrained, we tend to lose precise predictions for
the values of the gauge couplings in this setup. For instance, if we
introduce bulk masses for bulk fermion fields slightly smaller than
the compactification scale, which do not change the present analysis
of the effective potentials approximately, their contributions to the
evolution are suppressed.
The standard model fermions and the Higgs scalar 
can be introduced into the models as bulk fields or localized
fields in the present 
setup.\footnote{ 
This means that the hierarchy
problem is not addressed in the present study, as in usual
four-dimensional non-supersymmetric GUT models.}  As mentioned, the
models based on $SU(N)$ $(N=7,8)$ and $SO(10)$ share a part of
symmetry breaking pattern with the product GUT models~\cite{pgu} and
the flipped $SU(5)$ models~\cite{flipgut}, respectively, although the
construction of the hypercharge generator in our models is different
from the known models. This implies that the standard model fields are
realized as boundary localized fields in our $SU(N)$ and $SO(10)$
models. On the other hand, the standard model fields are naturally
incorporated into bulk fields in the $E_6$ model\footnote{ 
Recently,
the classification of the standard model fermions from bulk 27-plet
fields in models with orbifold breakings of the $E_6$ unified
symmetry is studied in Ref.~\cite{e6mat}.  }. 
In addition, the
supersymmetric extension of the $E_6$ model can supply the
doublet-triplet splitting via an analogues of the missing partner
mechanism that is often discussed in the flipped $SU(5)$ models. These
subjects are left to our future studies.

\bigskip
\subsection*{Acknowledgments}
The authors would like to thank N.~Yamatsu for valuable discussions.
\bigskip

\appendix
\section{Calculation of the effective potential in the $E_6$ model} 
\label{sec:apml}
\subsection{Contributions from bulk fermion fields} 
In Sec.~\ref{subsec:a5eff}, the contribution to the effective
potential from bulk fermion fields in the $E_6$ model is shown, with
the help of the effective potential in the $SO(10)$ model. In this
subsection, we derive the contribution by using another explicit
formulation.

For deriving the contribution, a key point is that the $U(1)$
directions accompanied by $\til d$ and $\til n$ in Eq.~\eqref{wdofe6} 
are identified to
generators in $SU(6)$ and $SU(2)_E$, respectively. This can be
realized because the zero mode of $A_5$ appears in the adjoint
representations of $SU(6)$ and $SU(2)_E$. To see this, we first focus
on the decomposition of $A_\mu$ in
Eqs.~\eqref{su5u1decomp1}--\eqref{su5u1decomp4}.  It is useful to
consider the $U(1)_{Y_F}$ subgroup that appears in
$SU(5)_F\supset SU(3)_C\times SU(2)_L\times U(1)_{Y_F}$. One can further
decompose the $SU(5)_F\times U(1)_{V_F}$ representations of $A_\mu$ as
\begin{align}
  \label{gsmdecomp1}
  A_\mu(24_0)^{(+,+)}&= A_\mu(G)^{(+,+)}+A_\mu(W)^{(+,+)}+A_\mu(n_{Y_F})^{(+,+)}
                       +A_\mu(Q)^{(+,+)}+A_\mu(\overline{Q})^{(+,+)},\\
  A_\mu(1^{V_F}_0)^{(+,+)}&=A_\mu(n_{V_F})^{(+,+)},\\
  A_\mu(10_{-4})^{(+,-)}&=A_\mu(X)^{(+,-)}                          +A_\mu(\overline{U})^{(+,-)}+A_\mu(E)^{(+,-)},\\
  A_\mu(\overline{10}_{4})^{(+,-)}&=A_\mu(\overline{X})^{(+,-)}                          +A_\mu({U})^{(+,-)}+A_\mu(\overline{E})^{(+,-)},\\
  A_\mu(1^{V'}_0)^{(+,+)}&=A_\mu(n_{V'})^{(+,+)},\\
  A_\mu(10_1)^{(-,-)}&=A_\mu(q)^{(-,-)}
                       +A_\mu(\overline{d})^{(-,-)}+A_\mu(\overline{n})^{(-,-)},\\
  A_\mu(\overline{5}_{-3})^{(-,+)}&=A_\mu(\overline{u})^{(-,+)}
                                    +A_\mu({\ell})^{(-,+)},\\
  A_\mu(1_5)^{(-,+)}&=A_\mu(\overline{e})^{(-,+)},\\
  A_\mu(\overline{10}_{-1})^{(-,-)}&=A_\mu(\overline{q})^{(-,-)}
                       +A_\mu({d})^{(-,-)}+A_\mu({n})^{(-,-)},\\
  A_\mu(5_{3})^{(-,+)}&=A_\mu({u})^{(-,+)}
                                    +A_\mu(\overline{\ell})^{(-,+)},\\
  A_\mu(1_{-5})^{(-,+)}&=A_\mu({e})^{(-,+)},
\label{gsmdecomp1end}
\end{align}
where $A_\mu(R)$ in the right-hand sides transforms as $R$ in
Table~\ref{3211table} under the
$SU(3)_C\times SU(2)_L\times U(1)_{Y_F}\times U(1)_{V_F}$ symmetry.
We denote the complex conjugate of $R$ by $\overline{R}$.  Note that
the $U(1)_Y$ hypercharge and the charge of the Cartan generator of
$SU(2)_R$, which we denote by $T^3_R$, are linear combinations of
$U(1)_{Y_F}$ and $U(1)_{V_F}$ charges; they are also shown in
Table~\ref{3211table}.  One can see that $SU(3)_C\times SU(2)_L$ gauge
fields correspond to the zero modes of $A_\mu(G)^{(+,+)}$ and
$A_\mu(W)^{(+,+)}$, and the $U(1)_Y$ gauge field is a linear
combination of the zero modes of $A_\mu(n_{Y_F})^{(+,+)}$ and
$A_\mu(n_{V_F})^{(+,+)}$.  If the symmetry breaking
$SU(5)_F\times U(1)_{V_F}\to G_{\rm SM}$ is realized, then the zero
modes $A_\mu(Q)^{(+,+)}$, $A_\mu(\overline{Q})^{(+,+)}$, and a linear
combination of $A_\mu(n_{Y_F})^{(+,+)}$ and $A_\mu(n_{V_F})^{(+,+)}$
become massive with would-be NG bosons that belong to the zero mode of
$A_5$.
\begin{table}
\begin{center}
\caption{Representations and charges 
under $SU(3)_C$, $SU(2)_L$, $U(1)_{Y_F}$, $U(1)_{V_F}$, $U(1)_{Y}$, 
and $T^3_{R}$. 
\bigskip
}
\label{3211table}
\bigskip
\begin{tabular}{cccccccccccccccc} 
\hline\hline
    & $G$ & $W$ & $n_{Y_F}$ & $Q$
    & $n_{V_F}$ & $U$ & $X$ & $E$ & $n_{V'}$ & $d$ & $q$ &$n$&$u$&$\ell$&$e$\\\hline
    $SU(3)_C$ & 8 & 1 & 1 &  3 & 1 & 3 & 3 & 1 & 1 & 3 & 3&1&3&1&1 \\
    $SU(2)_L$ & 1 & 3 & 1 & 2 & 1 & 1 & 2 &  1 & 1 & 1 & 2&1&1&1&1 \\
    $U(1)_{Y_F}$ & 0 & 0 & 0 & $-{5\over 6}$ & 0 & ${2\over 3}$ & ${1\over 6}$ &
    $1$ & 0 & ${2\over 3}$ & ${1\over 6}$&$-1$&$-{1\over 3}$&$-{1\over 2}$&$0$ \\
    $U(1)_{V_F} $& 0 & 0 & 0 & 0 & 0 & 4 & $-4$ &
    $-4$ & 0 & $-1$ & $1$&$-1$&$3$&$-3$&$-5$\\
    $U(1)_{Y} $& 0 & 0 & 0 & ${1\over 6}$  & 0 & ${2\over 3}$ & $-{5\over 6}$ &
    $-1$ & 0 & $-{1\over 3}$ & ${1\over 6}$&$0$&${2\over 3}$&$-{1\over 2}$&${-1}$\\
    $T^3_{R}$& 0 & 0 & 0 & $-{1\over 2}$  & 0 & 0 & ${1\over 2}$ &
    $1$ & 0 & ${1\over 2}$ & $0$&$-{1\over 2}$&$-{1\over 2}$&$0$&${1\over 2}$\\\hline\hline
  \end{tabular} 
\end{center}
\end{table}

We next focus on the $SU(6)_F\times SU(2)_{E_F}$ decomposition of
$A_5$.  Similarly to $A_\mu$ in
Eqs.~\eqref{su5decomp21}--\eqref{su5decomp23}, we can obtain
\begin{align}
  \label{su5decomp215}
 A_5((35,1))&=A_5(24_0)^{(-,-)}+A_5(5_{3})^{(+,-)}
                +A_5(\overline{5}_{-3})^{(+,-)}
                +A_5(1^{K_F}_0)^{(-,-)},\\
  A_5((1,3))&=A_5(1_5)^{(+,-)}+A_5(1_{-5})^{(+,-)}+A_5(1^{E_F}_0)^{(-,-)},\\
  A_5((20,2))&=A_5(10_1)^{(+,+)}+A_5(\overline{10}_{-1})^{(+,+)}
                 +A_5(10_{-4})^{(-,+)}+A_5(\overline{10}_{4})^{(-,+)}.
  \label{su5decomp235}
\end{align}
These equations can be rewritten by the representations in
Table~\ref{3211table} as
\begin{align}\notag
  A_5((35,1))= & A_5(G)^{(-,-)}+A_5(W)^{(-,-)}+A_5(n_{Y_F})^{(-,-)}
  +A_5(Q)^{(-,-)}+A_5(\overline{Q})^{(-,-)}\\
  &+A_5({u})^{(+,-)}+A_5(\overline{\ell})^{(+,-)} +
  A_5(\overline{u})^{(+,-)}  +A_5({\ell})^{(+,-)}+A_5({n_{K_F}})^{(-,-)},\\
  A_5((1,3))=& A_5(\overline{e})^{(+,-)}+ A_5({e})^{(+,-)}+
  A_5({n_{E_F}})^{(-,-)},\\
  \notag A_5((20,2))=& A_5(q)^{(+,+)}
  +A_5(\overline{d})^{(+,+)}+A_5(\overline{n})^{(+,+)}
  +A_5(\overline{q})^{(+,+)}+A_5({d})^{(+,+)}+A_5({n})^{(+,+)}\\
  &+A_5(X)^{(-,+)} +A_5(\overline{U})^{(-,+)}+
  A_5({E})^{(-,+)} + A_5(\overline{X})^{(-,+)}
  + A_5({U})^{(-,+)}+A_5(\overline{E})^{(-,+)}, 
\end{align}
where $A_5({n_{K_F}})^{(-,-)}$ and $A_5({n_{E_F}})^{(-,-)}$ are linear
combinations of $A_5(n_{V_F})^{(-,-)}$ and $A_5(n_{V'})^{(-,-)}$.
From the expression, using the $SU(2)_R$ flip, we can obtain the
$E_6\supset SU(6)\times SU(2)_E$ decomposition of $A_5$:
\begin{align}
A_5&=A_5((35,1)')+A_5((1,3)')+A_5((20,2)'), 
\end{align}
where the terms in the right-hand side transform as the irreducible
representations of $SU(6)\times SU(2)_E$. They are rewritten as
follows:
\begin{align}
\notag
  A_5((35,1)')= & A_5(G)^{(-,-)}+A_5(W)^{(-,-)}+A_5(n_{Y})^{(-,-)}
  +A_5(X)^{(-,+)} +A_5(\overline{X})^{(-,+)} \\\label{a5su6su2e1}
  & +A_5({d})^{(+,+)} +A_5(\overline{\ell})^{(+,-)}
  +A_5(\overline{d})^{(+,+)}+A_5({\ell})^{(+,-)}+A_5({n_K})^{(-,-)},\\
  A_5((1,3)')=& A_5(\overline{n})^{(+,+)}+A_5({n})^{(+,+)}+
  A_5({n_E})^{(-,-)},\\
  \notag A_5((20,2)')=& A_5(q)^{(+,+)} + A_5(\overline{u})^{(+,-)}+A_5(\overline{e})^{(+,-)}
  +A_5(\overline{q})^{(+,+)}+A_5({u})^{(+,-)}+A_5({e})^{(+,-)} \\
  &+A_5(Q)^{(-,-)}+A_5(\overline{U})^{(-,+)}+
  A_5(\overline{E})^{(-,+)} +A_5(\overline{Q})^{(-,-)}
  + A_5({U})^{(-,+)}+A_5({E})^{(-,+)}.
\label{a5su6su2e2}
\end{align}
In this expression, $A_5({n_{Y_F}})^{(-,-)}$, $A_5({n_{K_F}})^{(-,-)}$,
and $A_5({n_{E_F}})^{(-,-)}$ are rearranged as $A_5({n_{Y}})^{(-,-)}$,
$A_5({n_K})^{(-,-)}$, and $A_5({n_E})^{(-,-)}$.
One can now clearly see that $\til d$ and $\til n$ in
Eq.~\eqref{wdofe6}, which parametrize the zero mode of $A_5$ and thus
are involved in the parity $(+,+)$ fields in 
Eqs.~\eqref{a5su6su2e1}--\eqref{a5su6su2e2},
belong to the gauge field of an $SU(2)$ subgroup of $SU(6)$ and $SU(2)_E$,
respectively;
these $SU(2)$ symmetries are just what we called $SU(2)_d$ and $SU(2)_n$ 
below Eq.~\eqref{wdofe6}.

Let us derive the contributions to the effective potential for
$\til d$ and $\til n$ from a bulk adjoint field
$\Phi_{A}^{(\til \eta)}$, where $\til \eta$ is a parameter related to
the periodicity of the field. For this purpose, we show the
transformation law of $\Phi_{A}^{(\til \eta)}$ under
$SU(2)_d\times SU(2)_n$ accompanied by the periodicity of the field,
explicitly. The adjoint field is decomposed into $SU(6)\times SU(2)_E$
representations as
\begin{align}
\Phi_{A}^{(\til \eta)}&=\Phi_{A}^{(\til \eta)}((35,1)')
+\Phi_{A}^{(\til \eta)}((1,3)')
+\Phi_{A}^{(\til \eta)}((20,2)'). 
\end{align}
One can further decompose the fields as
\begin{align}
\notag
  \Phi_{A}^{(\til \eta)}((35,1)')= & \Phi_{A}(G)^{(+\til
    \eta)}+\Phi_{A}(W)^{(+\til \eta)}+\Phi_{A}(n_{Y})^{(+\til
    \eta)}
  +\Phi_{A}(X)^{(-\til \eta)} +\Phi_{A}(\overline{X})^{(-\til \eta)} \\\label{apdecomp11}
  & +\Phi_{A}({d})^{(+\til \eta)} +\Phi_{A}(\overline{\ell})^{(-\til
    \eta)} +\Phi_{A}(\overline{d})^{(+\til
    \eta)}+\Phi_{A}({\ell})^{(-\til \eta)}
  +\Phi_{A}({n_K})^{(+\til \eta)},\\\label{apdecomp12}
  \Phi_{A}^{(\til \eta)}((1,3)')=& \Phi_{A}(\overline{n})^{(+\til
    \eta)}+\Phi_{A}({n})^{(+\til \eta)}+
  \Phi_{A}({n_E})^{(+\til \eta)},\\
  \notag \Phi_{A}^{(\til \eta)}((20,2)')=& \Phi_{A}(q)^{(+\til \eta)} + \Phi_{A}(\overline{u})^{(-\til
    \eta)}+\Phi_{A}(\overline{e})^{(-\til \eta)}
  +\Phi_{A}(\overline{q})^{(+\til \eta)}+\Phi_{A}({u})^{(-\til
    \eta)}+\Phi_{A}({e})^{(-\til \eta)}\\
  &+\Phi_{A}(Q)^{(+\til
    \eta)}+\Phi_{A}(\overline{U})^{(-\til \eta)}+
  \Phi_{A}(\overline{E})^{(-\til \eta)}
  +\Phi_{A}(\overline{Q})^{(+\til \eta)}
  + \Phi_{A}({U})^{(-\til \eta)}+\Phi_{A}({E})^{(-\til \eta)}, \label{apdecomp13}
\end{align}
where in the right-hand sides $\pm \til \eta$ denotes 
the periodicity that coincides with the
eigenvalue of the translation operator $\hat P_1 \hat P_0$. 

From the above expression, we can easily see the $SU(2)_n$
transformation law of $\Phi_{A}^{(\til \eta)}$ with the periodicity.
In $\Phi_{A}^{(\til \eta)}((1,3)')$, there is one triplet of the
periodicity $+ \til \eta$. In $\Phi_{A}^{(\til \eta)}((20,2)')$, there
are twenty doublets, of which the twelve have $+ \til \eta$ and the
rest have $- \til \eta$.

For $SU(2)_d$, a little difficulty remains; the transformation law is
not manifest in Eqs.~\eqref{apdecomp11}--\eqref{apdecomp13} since
$SU(2)_d$ does not commute with $SU(3)_C$. Fortunately, we can find
another $SU(2)$ that helps us to see the $SU(2)_d$ transformation law
of each $SU(6)\times SU(2)_E$ representation in
Eqs.~\eqref{apdecomp11}--\eqref{apdecomp13} taking account of the
periodicity $\pm \til \eta$.  Note that the $SU(4)$ subgroup in
$SU(6)/SU(2)_L$ commutes with the translation operator
$\hat P_1 \hat P_0$ and involves both $SU(3)_C$ and $SU(2)_d$.  This
implies that using any $SU(2)$ subgroup of $SU(3)_C$ instead of
$SU(2)_d$, we can derive the $SU(2)_d$ transformation law of each
$SU(6)\times SU(2)_E$ representation with the
periodicity. Accordingly, under $SU(2)_d$, it is realized that there
are one triplet of $+ \til \eta$, four doublets of $+ \til \eta$, and
four doublets of $- \til \eta$ in $\Phi_{A}^{(\til \eta)}((35,1)')$.
For $\Phi_{A}^{(\til \eta)}((20,2)')$, eight doublets of $+ \til \eta$
and four doublets of $- \til \eta$ are contained. Among the doublets,
since $\Phi_{A}^{(\til \eta)}((20,2)')$ is also $SU(2)_n$ doublet,
there are bi-doublets under $SU(2)_d\times SU(2)_n$; four bi-doublets
of $+ \til \eta$ and two bi-doublets of $- \til \eta$ are contained.

In this way, the decompositions in
Eqs.~\eqref{apdecomp11}--\eqref{apdecomp13} tell us the
transformations of the fields under $SU(2)_d\times SU(2)_n$.  The
result is
\begin{align}
  \Phi_{A}^{(\til \eta)}((35,1)')&\ni 
4\times\Phi_{A}(2,1)^{(+\til \eta)} +
4\times \Phi_{A}(2,1)^{(-\til \eta)}
+1\times \Phi_{A}(3,1)^{(+\til \eta)},\\
  \Phi_{A}^{(\til \eta)}((1,3)')&\ni 
1\times \Phi_{A}(1,3)^{(+\til \eta)},\\
  \Phi_{A}^{(\til \eta)}((20,2)')&\ni 4\times 
\Phi_{A}(1,2)^{(+\til \eta)} +4\times
  \Phi_{A}(1,2)^{(-\til \eta)} +4\times \Phi_{A}(2,2)^{(+\til \eta)}
+ 2\times \Phi_{A}(2,2)^{(-\til \eta)}, 
\end{align}
where in the right-hand sides $\Phi_{A}(R,R')$ transforms under
$SU(2)_d$ $(SU(2)_n)$ as $R$ $(R')$ and the superscript for each term 
denotes the periodicity. Hence one can obtain
\begin{align}\notag
\Phi_{A}^{(\pm)}\ni ~
&4 \left[ \Phi_{A}(2,1)^{(\pm)}+\Phi_{A}(2,1)^{(\mp)}+ \Phi_{A}(1,2)^{(\pm)}+\Phi_{A}(1,2)^{(\mp)} \right]
\\&  +4\Phi_{A}(2,2)^{(\pm)}+2\Phi_{A}(2,2)^{(\mp)}
+\Phi_{A}(3,1)^{(\pm)}+\Phi_{A}(1,3)^{(\pm)}.
\label{adjdecompap}
\end{align}
From the above, the contribution in Eq.~\eqref{potf78n} is obtained. 

The contribution from a 27-plet $\Phi_{F}^{(\til \eta)}$ is obtained 
in a similar fashion. The field is decomposed into 
$SU(6)_F\times SU(2)_{E_F}$ multiplets as 
\begin{align}
  \Phi_{F}^{(\til \eta)}&=\Phi_{F}^{(\til \eta)}((15,1))+\Phi_{F}^{(\til \eta)}
((\overline{6},2)). 
\end{align}
Further decomposition leads to 
\begin{align}
  \Phi_{F}^{(\til \eta)}((15,1))&=
\Phi_{F}(q)^{(-\til \eta)}+ 
\Phi_{F}(\overline{d})^{(-\til \eta)}+ 
\Phi_{F}(\overline{n})^{(-\til \eta)}+ 
\Phi_{F}(D)^{(+\til \eta)}+ 
\Phi_{F}(L)^{(+\til \eta)},\\
  \Phi_{F}^{(\til \eta)}((\overline{6},2))&=
\Phi_{F}(\overline{u})^{(+\til \eta)}+ 
\Phi_{F}(\ell)^{(+\til \eta)}+ 
\Phi_{F}(s)^{(-\til \eta)}+ 
\Phi_{F}(\overline{D})^{(-\til \eta)}+ 
\Phi_{F}(\overline{L})^{(-\til \eta)}+ 
\Phi_{F}(\overline{e})^{(+\til \eta)}.
\end{align}
In the right-hand sides $\Phi_{F}(R)$ transforms under
$SU(3)_C\times SU(2)_L\times U(1)_{Y_F}\times U(1)_{V_F}$ symmetry as
in Table~\ref{3211table} and~\ref{3211table2} and the superscript for
each term denotes the periodicity.
\begin{table}
\begin{center}
\caption{Representations and charges 
under $SU(3)_C$, $SU(2)_L$, $U(1)_{Y_F}$, $U(1)_{V_F}$, $U(1)_{Y}$, 
and $T^3_{R}$. 
\bigskip
}
\label{3211table2}
\bigskip
\begin{tabular}{cccc}
\hline \hline 
    & $D$ & $L$ & $s$ \\\hline
    $SU(3)_C$ & 3 & 1 & 1 \\
    $SU(2)_L$ & 1 & 2 & 1 \\
    $U(1)_{Y_F}$ & $-{1\over 3}$ & ${1\over 2}$ & 0 \\
    $U(1)_{V_F} $& $-2$ & $-2$ & 0 \\
    $U(1)_{Y} $& $-{1\over 3}$ & $-{1\over 2}$ & 0\\
    $T_{3R} $& $0$ & ${1\over 2}$ & 0 \\\hline\hline
  \end{tabular} 
\end{center}
\end{table}
Using the $SU(2)_R$ flip, we can obtain the
$E_6\supset SU(6)\times SU(2)_E$ decomposition as
\begin{align}
  \Phi_{F}^{(\til \eta)}((15,1)')&=
\Phi_{F}(q)^{(-\til \eta)}+ 
\Phi_{F}(\overline{u})^{(+\til \eta)}+ 
\Phi_{F}(\overline{e})^{(+\til \eta)}+ 
\Phi_{F}(D)^{(+\til \eta)}+ 
\Phi_{F}(\overline{L})^{(-\til \eta)}
,\\
  \Phi_{F}^{(\til \eta)}((\overline{6},2)')
&=
\Phi_{F}(\overline{d})^{(-\til \eta)}+ 
\Phi_{F}(\ell)^{(+\til \eta)}+ 
\Phi_{F}(s)^{(-\til \eta)}+ 
\Phi_{F}(\overline{D})^{(-\til \eta)}+ 
\Phi_{F}(L)^{(+\til \eta)}+ 
\Phi_{F}(\overline{n})^{(-\til \eta)}.
\end{align}
Therefore, a 27-plet involves 
the fields that transform under $SU(2)_d\times SU(2)_n$ as 
\begin{align}
\Phi_{F}^{(\til \eta)}((15,1)')&\ni
2\times \Phi_{F}(2,1)^{(+\til \eta)}+
2\times \Phi_{F}(2,1)^{(-\til \eta)},\\
\Phi_{F}^{(\til \eta)}((\overline{6},2)')&\ni 
2\times \Phi_{F}(1,2)^{(+\til \eta)}+
2\times \Phi_{F}(1,2)^{(-\til \eta)}+
1\times \Phi_{F}(2,2)^{(-\til \eta)}.
\end{align}
Thus one can realize 
\begin{align}\notag
\Phi_{F}^{(\pm)}&\ni
2\left[\Phi_{F}(2,1)^{(\pm)}+\Phi_{F}(2,1)^{(\mp)}
+\Phi_{F}(1,2)^{(\pm)}+\Phi_{F}(1,2)^{(\mp)}\right]
+\Phi_{F}(2,2)^{(\mp)}.
\end{align}
From the above, we can easily lead to Eq.~\eqref{potf27p}.

\subsection{Contributions from the gauge field} 
\label{sec:apanom}
In this subsection, we show the calculation of the contributions to
the effective potential discussed in Sec.~\ref{subsec:effanom}.  The
contribution is generated by the gauge field, whose $U(1)_{V'}$
component is assumed to have a large mass term at the $y=0$ boundary
due to the anomaly cancellation. The mass term effectively modifies
the boundary condition and the KK masses of some components of the
gauge field.  Without the boundary mass term, the contribution takes
the form of Eq.~\eqref{potf78n} with $\delta=0$, since the gauge field
belongs to the adjoint representation. The modification of the KK mass
alters a part of the contribution in Eq.~\eqref{potf78n}.

As explained in Sec.~\ref{subsec:effanom}, the KK mass spectrum that is
affected by the localized mass term is obtained as a solution to the
EOM of the following set of the fields:
\begin{align}\label{anomamu}
\left(A_\mu(n^{(3)}), A_\mu(n^{(2)}), A_\mu(d^{(3)}), A_\mu(d^{(2)}), A_\mu(X)\right), 
\end{align}
where $n^{(3,2)}$, $d^{(3,2)}$, and $X$ imply generators of $SU(2)_n$,
$SU(2)_d$, and $U(1)_X$, respectively. For convenience we introduce a
column vector $\Phi_\mu^{\alpha}$ $(\alpha=1\textrm{--}5)$ that
consists of the fields:
\begin{align}
 \Phi_\mu^{\alpha} \equiv 
\left({A_\mu(d^2)-iA_\mu (d^3)\over \sqrt{2}}, 
{A_\mu(d^3)-iA_\mu (d^2)\over \sqrt{2}}, 
A_\mu(X),
{A_\mu(n^3)-iA_\mu (n^2)\over \sqrt{2}},
{A_\mu(n^2)-iA_\mu (n^3)\over \sqrt{2}}
\right)^{ T}.
\end{align}
The Lagrangian in Eq.~\eqref{laggauge} is diagonalized by 
the vector as 
\begin{align}
  {\cal L}_{\rm gauge}&\ni 
{1\over 2}\eta^{\mu\nu}(\Phi_\mu^{\alpha})^\dag
\left[\square \delta^{\alpha\beta}-\til{\cal
    D}^{\alpha\beta}\right]\Phi_\nu^\beta, \qquad 
\delta^{\alpha\beta}={\rm diag}(1,1,1,1,1),\\
\til{\cal D}^{\alpha\beta}&={\rm diag} \left( (\partial_5-i{\til d/
    R})^2, (\partial_5+i{\til d/ R})^2, (\partial_5)^2, 
(\partial_5+i{\til n/ R})^2, 
(\partial_5-i{\til n/ R})^2 \right).
\end{align}
We introduce the KK mode expansion:
\begin{align}
  \Phi_\mu^\alpha(x,y)=\sum_{n=-\infty}^\infty\psi_\mu^{(n)}(x)
\phi^{\alpha(n)}(y), \qquad (\square+m_{\rm KK}^{(n)2})\psi_\mu^{(n)}(x)=0. 
\end{align}
The solution to the bulk EOM is obtained as follows: 
\begin{align}
\label{anomsol1}
  \phi^{1(n)}(y)&=
  e^{i\til d y/R}\left[\xi^1 \cos(m_{\rm KK}^{(n)}y)+\zeta^1 \sin(m_{\rm KK}^{(n)}y)\right],\\
  \phi^{2(n)}(y)&=
  e^{-i\til d y/R}\left[\xi^2 \cos(m_{\rm KK}^{(n)}y)+\zeta^2 \sin(m_{\rm KK}^{(n)}y)\right],\\
  \phi^{3(n)}(y)&=
  \xi^3 \cos(m_{\rm KK}^{(n)}y)+\zeta^3 \sin(m_{\rm KK}^{(n)}y),\\
  \phi^{4(n)}(y)&=
  e^{-i\til n y/R}\left[\xi^4 \cos(m_{\rm KK}^{(n)}y)+\zeta^4 \sin(m_{\rm KK}^{(n)}y)\right],\\
  \phi^{5(n)}(y)&= 
e^{i\til n y/R}\left[\xi^5 \cos(m_{\rm KK}^{(n)}y)+\zeta^5 \sin(m_{\rm KK}^{(n)}y)\right], 
\label{anomsol5}
\end{align}
where $\xi^\alpha$ and $\zeta^\alpha$ $(\alpha=1\textrm{--} 5)$ are
real constants and $m_{\rm KK}^{(n)}$ is the mass of the $n$-th KK mode.

In order to determine the KK mass $m_{\rm KK}^{(n)}$, the
boundary condition at $y=0$ and $\pi R$ should be imposed to the
solutions in Eqs.~\eqref{anomsol1}--\eqref{anomsol5}. In the present
case, the condition is simplified in a basis where $U(1)_{V'}$ is
manifest, since there is a boundary mass term for $A_\mu(V')$.  We
introduce the new basis,
\begin{align}
  \Phi_\mu'^{\alpha}
  =U^{\alpha\beta}\Phi_\mu^{\beta}, \qquad 
  U^{\alpha\beta}=
{1\over 4\sqrt{5}}
   \begin{pmatrix}
2\sqrt{10}&2i\sqrt{10}&0&0&0\\
4 i&4&-4\sqrt{3}&0&0\\
i \sqrt{15}&\sqrt{15}&2\sqrt{5}&\sqrt{15}&i\sqrt{15}\\
3i&3&2\sqrt{3}&-5&-5i\\
0&0&0&2i\sqrt{10}&2\sqrt{10}
   \end{pmatrix}, 
\end{align}
and we denote
\begin{align}
  \Phi_\mu'^{\alpha}&= (A_\mu(d^2),A_\mu(K'),A_\mu(V'),A_\mu(V),A_\mu(n^2))^{ T}, 
\end{align}
where $A_\mu(K')$, $A_\mu(V')$, and $A_\mu(V)$ are defined by
\begin{align}
  \begin{pmatrix}
    A_\mu(K')\\A_\mu(K)
  \end{pmatrix}&=
{1\over \sqrt{5}}
  \begin{pmatrix}
    {\sqrt{2}}&{-\sqrt{3}}\\
    {\sqrt{3}}&{\sqrt{2}}
  \end{pmatrix}
  \begin{pmatrix}
    A_\mu(d^3)\\A_\mu(X)
  \end{pmatrix},&
  \begin{pmatrix}
    A_\mu(V')\\A_\mu(V)
  \end{pmatrix}&=
{1\over 4}  \begin{pmatrix}
    {\sqrt{10}}&{\sqrt{6}}\\
    {\sqrt{6}}&-{\sqrt{10}}
  \end{pmatrix}
  \begin{pmatrix}
    A_\mu(K)\\A_\mu(n^3)
  \end{pmatrix}. 
\end{align}
It is realized that $A_\mu(V')$ and $A_\mu(V)$ correspond to the 
$U(1)_{V'}$ and $U(1)_{V}$ gauge field, respectively. 
By using the above fields, the boundary condition is simplified; 
at $y=0$, the condition is written as 
\begin{align}
  & \Phi_{\mu}'^{1}=\Phi_{\mu}'^{5}=0, \qquad 
\del_y\Phi_{\mu}'^{2}=\del_y\Phi_{\mu}'^{4}=0, \qquad 
(2\del_y-M)\Phi_{\mu}'^3=0, 
\label{modbc1}
\end{align}
where $M$ represents the boundary mass parameter.  On the other
hand, at $y=\pi R$, the condition is written as
\begin{align}
  \Phi_{\mu}'^{1}=\Phi_{\mu}'^{5}=0,\qquad 
  \del_y\Phi_{\mu}'^{2}= \del_y\Phi_{\mu}'^{3}= \del_y\Phi_{\mu}'^{4}
  =0. 
\label{modbc2}
\end{align}

Imposing the boundary condition to the solutions in
Eqs.~\eqref{anomsol1}--\eqref{anomsol5}, we can determine the KK mass
$m_{\rm KK}^{(n)}$. In our model, we are interested in the case where
$M$ is much larger than the compactification scale $1/R$. In this
case, we obtain
\begin{align}
    \sin^2(m_{\rm KK}^{(n)}\pi R)&={S_1\pm S_2\over 8}, 
\end{align}
where $S_1$ and $S_2$ are found in Eq.~\eqref{s1s2}. The above
equation leads to Eqs.~\eqref{mkk1} and~\eqref{mkk2}.

If there were no boundary mass term, the gauge field and a bulk
adjoint field with $\delta=0$, which gives the contribution
$F^{78}(\til d,\til n,0)$ in Eq.~\eqref{potf78n}, have the same KK
mass spectrum. In this case, the $n$-th KK masses
of the fields in Eq.~\eqref{anomamu} are 
\begin{align}
\bigg({n\pm \til d\over R}\bigg)^2,\quad 
\bigg({n\pm \til n\over R}\bigg)^2, 
\label{kkmassnblimit}
\end{align}
and ${n^2/R^2}$. In the contribution $F^{78}(\til d,\til n,0)$, the KK
masses in Eq.~\eqref{kkmassnblimit} are related to the terms
proportional to $\hat f(2\til d)$ and $\hat f(2\til n)$. The boundary
mass term alter the KK masses in Eq.~\eqref{kkmassnblimit} into the
form in Eqs.~\eqref{mkk1} and~\eqref{mkk2}. Thus the contribution from
the gauge field with a large boundary mass is obtained by the
replacement $\hat f(2\til d) $ and $\hat f(2\til n)$ with
$\hat f(2\rho_+) $ and $\hat f(2\rho_-) $ in
$F^{78}(\til d,\til n,0)$.  The result is shown in Eq.~\eqref{fa}.

While the above discussions focus on the contributions from $A_\mu$,
we also incorporate the contributions from $A_5$. 
Although the $U(1)_{V'}$ component of $A_5$ has neither zero mode nor
direct coupling to the localized mass, the boundary condition of the
field is modified from the Dirichlet to the Neumann type effectively by a
large boundary mass ($M\gg 1/R$). This modification is
induced by proper treatment of gauge fixing terms. In
ref.~\cite{A5KK}, a similar situation can be seen in terms of the
four-dimensional effective description.  In Appendix~\ref{sec:bcmod},
five-dimensional treatment in a simple $U(1)$ case is shown as an
illustrative example.\footnote{Also in ref.~\cite{A55d}, discussion
  about five-dimensional treatment is found, while the form of the
  gauge fixing terms are different from our example.}

\bigskip
\section{Equivalence of vacua in the $E_6$ model} 
\label{sec:mod}
As mentioned in Sec.~\ref{subsec:a5eff}, the effective potential in
the $E_6$ model has the invariance and the periodicity under the
characteristic transformation
$(\til n,\til d)\to (\til n+1,\til d+1)$.  In this section, we show
the invariance of the potential is ensured by the $E_6$ gauge
transformation.

In a five-dimensional orbifold model, the gauge transformation is
generally written by
\begin{align}
\label{gengaugetrans}
A_M(x,y) &\to  A_M'(x,y)=\Lambda(x,y)A_M(x,y)\Lambda^\dag(x,y) +{i\over g}
\Lambda(x,y)\del_M\Lambda^\dag(x,y), 
\end{align}
where
\begin{align}
 \Lambda(x,y)&=\exp\left({i\lambda_a(x,y)t^a}\right), 
\end{align}
and $\lambda_a(x,y)$ is a five-dimensional gauge transformation
function. When the gauge field satisfies the boundary condition given
in Eqs.~\eqref{bcsgen1}--\eqref{bcsgen2}, then the gauge transformation
implies
\begin{align}
\label{bcmod1}
A_\mu'(x,-y)  &=P_0'A_\mu'(x,y)P_0'^\dag+{i\over g}P_0'\del_\mu P_0'^\dag,
\quad 
A_\mu'(x,\pi R-y)  =P_1'A_\mu'(x,\pi R+y)P_1'^\dag+{i\over g}P_1'\del_\mu P_1'^\dag,\\
\label{bcmod4}
A_5'(x,-y)  &=-P_0'A_5'(x,y)P_0'^\dag-{i\over g}P_0'\del_5 P_0'^\dag,\quad 
A_5'(x,\pi R-y)  =-P_1'A_5'(x,\pi R+y)P_1'^\dag-{i\over g}P_1'\del_5 P_1'^\dag,
\end{align}
where 
\begin{align}
  P_0'=\Lambda(x,-y)P_0\Lambda^\dag(x,y), \qquad 
  P_1'=\Lambda(x,\pi R-y)P_1\Lambda^\dag(x,\pi R+y).
\end{align}

Although generally the gauge transformation changes the boundary
conditions, one can find the particular gauge transformation such that
the relations $P_0=P_0'$ and $P_1=P_1'$ hold and the gauge field is
shifted as $A'_M\neq A_M$. In this case, the non-linear terms in
Eqs.~\eqref{bcmod1} and~\eqref{bcmod4} vanish and hence the same
boundary conditions are imposed on $A_M$ and $A'_M$. For example,
suppose that a vacuum configuration $\vev{A_5}$ is shifted to
$\vev{A'_5}$ by the gauge transformation function that preserves
the boundary conditions, then the potential should have degenerate vacua
around $\vev{A_5}$ and $\vev{A_5}'$.

We start to discuss our $E_6$ model, where the VEV 
in Eq.~\eqref{wdofe6} can be written as 
\begin{align}
  \vev{A_5}&={1\over gR}(\tilde d t_{\tilde d}+\tilde n t_{\tilde n}), 
\label{a5paraap1}
\end{align}
where $t_{\tilde d}$ and $t_{\tilde n}$ are generators of
$SU(2)_{d}\subset SU(6)$ and $SU(2)_{n}=SU(2)_E$, respectively.  In a
definite basis of the fundamental representation of
$SU(6)\times SU(2)_E$, the parity matrix of the boundary condition can
be written as follows:
\begin{align}
\label{p0rep}
  P_0&={\rm diag}(+1,+1,+1,+1,+1,-1)\otimes{\rm diag}(+1,-1),\\
\label{p1rep}
  P_1&={\rm diag}(+1,+1,+1,-1,-1,-1)\otimes{\rm diag}(+1,-1). 
\end{align}
Here we take that $t_{\tilde d}$ generates 
mixing between 1st and 6th entries in the $SU(6)$ fundamental representation. 
As discussed in Sec.~\ref{sec:general}, the generators and the parity matrix 
satisfy
\begin{align}
  \{P_0,t_{\tilde d}\}=\{P_0,t_{\tilde n}\}=\{P_1,t_{\tilde d}\}=\{P_1,t_{\tilde n}\}=0, \qquad 
[t_{\tilde d},t_{\tilde n}]=0.
\end{align}

Let us now consider the gauge transformation 
\begin{align}\label{apgt}
  \Lambda(x,y)&=\exp\left({i{y\over R}(t_{\tilde d}+t_{\tilde n})}\right).
\end{align}
From Eq.~\eqref{gengaugetrans}, one can see that the transformation
shifts the parameters in Eq.~\eqref{a5paraap1} as
\begin{align}\label{apvevs}
  (\til d,\til n)\to (\til d',\til n')=  (\til d+1,\til n+1).
\end{align}
The gauge-transformed field satisfies the boundary condition in
Eq.~\eqref{bcmod4} with the new parity matrices
\begin{align}
  P_0'&=P_0, \qquad P_1'=\exp\left({2\pi i t_{\tilde d}}\right)\exp\left({2\pi i t_{\tilde n}}\right)P_1. 
\end{align}
As mentioned above, if $P_1'=P_1$ is satisfied, then the effective
potential should have invariance under the shift in
Eq.~\eqref{apvevs}. This can be shown as follows.  In the
$SU(6)\times SU(2)_E$ representation space in Eqs.~\eqref{p0rep}
and~\eqref{p1rep}, the matrices $\exp({2\pi i t_{\tilde d}})$ and
$\exp({2\pi i t_{\tilde n}})$ correspond to $2\pi$ rotations of
fundamental representation of $SU(2)_d$ and $SU(2)_E$,
respectively. Thus we obtain
\begin{align}
\exp\left({2\pi i t_{\tilde d}}\right)&=  {\rm diag}(-1,+1,+1,+1,+1,-1)\otimes {\rm diag}(+1,+1),\\ 
\exp\left({2\pi i t_{\tilde n}}\right)&={\rm diag}(+1,+1,+1,+1,+1,+1)\otimes 
{\rm diag}(-1,-1). 
\end{align}
The parity matrix $P_1'$ explicitly written as follows: 
\begin{align}
P_1'&={\rm diag}(-1,+1,+1,-1,-1,+1)\otimes
{\rm diag}(-1,+1)
\notag \\
&={\rm diag}(+1,-1,-1,+1,+1,-1)\otimes
{\rm diag}(+1,-1)\\
&\to {\rm diag}(+1,+1,+1,-1,-1,-1)\otimes
{\rm diag}(+1,-1)=P_1. 
\end{align}
In the last line, we use an $SU(4)\subset SU(6)/SU(2)_d$ rotation;
this rotation does not change $P_0$ and the VEV since both $P_0$ and
$t_{\tilde d}$ commute with the $SU(4)$ subgroup. Although the
boundary condition in the $E_6$ model is unchanged under the gauge
transformation in Eq.~\eqref{apgt}, the transformation shifts the
parameters as in Eq.~\eqref{apvevs}. Therefore, the effective potential
should be invariant against the shift in Eq.~\eqref{apvevs}. This
leads to the periodicity in the potential.

\bigskip
\section{Effective modification of the boundary condition of $A_5$
  with boundary breaking}
\label{sec:bcmod}
In Sec.~\ref{sec:apanom}, we show that the boundary condition of
$A_\mu$ is effectively modified due to the existence of the boundary
mass term. As mentioned, while the $U(1)_{V'}$ component of $A_5$ does not
have zero modes, one can see that the boundary condition of $A_5$ is
also modified by introducing proper gauge fixing terms in the
theory. As a result, one can choose the specific gauge, namely $\xi=1$
shown just below, where the KK masses of $A_5$ coincide with those of
$A_\mu$. Here, we consider a five-dimensional $U(1)$ model
compactified on an $S^1/{\mathbb Z}_2$ orbifold, and illustrate the
essential feature of the modification in the simple setup.

In the $U(1)$ model, the orbifold parity around the fixed point $y=0$
is expressed as
\begin{align}
 A_\mu(-y)=A_\mu(y),\qquad A_5(-y)=-A_5(y).
\end{align}
Then, as the $E_6$ model discussed in Sec.~\ref{sec:e6}, we study
the effect of a mass term localized on this fixed point.  Below, we
consider an anomaly as the origin of the mass term, while it can be
the Higgs mechanism.

The anomaly is assumed to be made harmless via the Green-Schwarz
mechanism~\cite{GS}.  Namely, a pseudo-scalar field $\chi$ that
transforms non-linearly under the $U(1)$ symmetry and has the
Wess-Zumino couplings is introduced on this boundary to cancel the
anomaly.  Such a scalar field allows the St\"{u}ckelberg mass
term~\cite{Stuckelberg} (on the boundary)
\begin{align}
{\cal L}_{\rm St}={1\over 2} M\left( A_\mu+{1\over \sqrt{M}}\del_\mu\chi\right)^2\delta(y), 
\end{align}
which is $U(1)$ invariant and thus the naive scale of the mass $M$ is
around the cutoff scale of the five-dimensional theory, much larger
than the compactification scale.  As well-known, such a huge mass
repels the wave functions of the lower-laying KK modes of $A_\mu$
to modify its boundary condition from the Neumann to the Dirichlet
type effectively~\cite{mixedbc}. Below, we examine the effect on those
of $A_5$, which does not directly couple to the localized mass due
to the orbifold parity.  (See Ref.~\cite{A5KK} for the same analysis
in terms of the KK decomposed language.)

For this purpose, as $A_5$ is unphysical except for the zero mode, we
should treat the gauge fixing term properly.\footnote{ This means
  that, of course, the effect is gauge dependent and thus an unusual
  gauge fixing term may be selected as in Ref.~\cite{iGUT} to make the
  "mass spectrum" of $A_5$ unchanged.  In such cases, however, the
  calculation of, for instance, the effective potential would be
  complicated.}  The mixing terms of the four-dimensional gauge field
are
\begin{align}
{\cal L}_{\rm mix} = -\del_5A^\mu\del_\mu A_5 + \sqrt MA^\mu\del_\mu\chi\delta(y), 
\end{align}
and we adopt the usual gauge fixing term with a constant gauge
parameter $\xi$,
\begin{align} {\cal L}_{\rm GF} = -{1\over {2\xi}}\left[\del_\mu A^\mu
        - \xi\left(\del_5A_5+\sqrt M\chi\delta(y)\right) \right]^2,
\end{align}
to remove the above mixing terms (up to surface terms).  Then the
quadratic terms of $A_5$ and $\chi$ become 
\begin{align} {\cal L}_{\rm quad} = 
{1\over 2}A_5\left(-\square+\xi\del_5^2\right) A_5
    -\delta(y)\left(\xi\sqrt M\chi\del_5A_5+{1\over
          2}\chi\left(\square+\xi M\delta(y)\right)\chi\right).
\end{align}
Note that this quadratic part is essentially the same as the one in
the case that the mass term originates from the Higgs mechanism, and
thus the derivation below is applied also for the case.

Since there is an awkward term proportional to $\delta(y)^2$ in the quadratic
part, we should regularize the delta function.  To be more concrete,
we replace the delta function by a finite, sufficiently smooth
function $\delta_\epsilon(y)$ that vanishes for
$\left|y\right|\geq\epsilon$ and is normalized as
$\int_{-\epsilon}^\epsilon\delta_\epsilon(y)dy\sim1$.\footnote{ 
One may impose the periodicity, for completeness if necessary.} 
The EOMs of $A_5$ and $\chi$ are respectively
\begin{align}
& \left(-\square+\xi\del_5^2\right) A_5(y)+\xi\left(\del_5\delta_\epsilon(y)\right)\sqrt M\chi=0, 
\label{A5EOM}\\
& -\delta_\epsilon(y)\left(\xi\sqrt M\del_5A_5(y)
    +\left(\square+\xi M\delta_\epsilon(y)\right)\chi\right)=0, 
\label{chiEOM}
\end{align}
where the $y$-dependences are explicitly shown.  Due to the overall
delta function in Eq.~\eqref{chiEOM}, we may not suppose that the
combination in the parenthesis there vanishes for
$\left|y\right|\geq\epsilon$, while it does vanish, for instance, at
$y=0$ as
\begin{align}
  \xi\sqrt M\del_5A_5(0) + \left(\square+\xi M\delta_\epsilon(0)\right)\chi=0. 
\label{chiEOM0}
\end{align}
As often done, we integrate Eq.~\eqref{A5EOM}, which is an odd
function of $y$, over a tiny region
$0\leq y\leq z = \cal O(\epsilon)$.  Then, the contribution of the
regular function, $-\square A_5$ is negligible and we get
\begin{align}
 \xi\left[\del_5A_5(y) + \delta_\epsilon(y)\sqrt M\chi\right]^z_0 = 0.
\end{align}
Using Eq.~\eqref{chiEOM0} to remove the factor $\delta_\epsilon(0)$,
we obtain
\begin{align}
 \xi\left(\del_5A_5(z) + \delta_\epsilon(z)\sqrt M\chi\right)+\frac1{\sqrt M}\square \chi=0. 
\label{EOM2}
\end{align}
Its integration over again a tiny region,
$-\epsilon\leq z\leq\epsilon$, leads to 
\begin{align}
 \xi\left(2A_5(\epsilon)+\sqrt M\chi\right) = 0,
\label{EOM3}
\end{align}
where we use $A_5(-\epsilon)=-A_5(\epsilon)$.  Operating the
four-dimensional Laplacian $\square$ on Eq.~\eqref{EOM3} and applying
Eq.~\eqref{EOM2} evaluated at $y=\epsilon$ where the delta function
vanishes, we can derive an effective mixed boundary condition
\begin{align}
 2\xi\square A_5(\epsilon)-\xi^2 M\del_5A_5(\epsilon)=0.
\end{align}
The result shows that $A_5$ obeys the Dirichlet boundary condition in
the limit $M\to0$; the condition changes to the Neumann boundary
condition in the opposite limit $M\to\infty$. The modification of the
boundary condition of $A_5$ is in accordance with that of $A_\mu$.

\bigskip

\end{document}